\documentclass{PoS}
\usepackage{graphicx}

\title{Mechanisms of astrophysical jet formation, and comparison with laboratory experiments}

\ShortTitle{Astrophysical jet formation}

\author{\speaker{G.~S.~Bisnovatyi-Kogan}
\\
        Space Research Institute,  Russian Academy of Sciences,\\ Profsoyuznaya 84/32, Moscow 117997, Russia\\ and \\National Research Nuclear University MEPhI \\ (Moscow Engineering Physics Institute), Kashirskoe Shosse 31, Moscow 115409, Russia\\and\\
        Moscow Institute of Physics and Technology MIPT, Dolgoprudny, Moscow reg., Russia\\
        E-mail: \email{gkogan@iki.rssi.ru}}

\abstract{Jets are observed in young stellar objects, X-ray sources, active galactic nuclei (AGN). The mechanisms of jet formation may be divided in regular, acting continuously for a long time, and explosive ones [1]. Continuous mechanisms are related with electrodynamics and radiation pressure acceleration, hydrodynamical acceleration in the nozzle inside a thick disk, acceleration by relativistic beam of particles. Explosive jet formation is connected with supernovae, gamma ray bursts and explosive events in galactic nuclei. Mechanisms of jet collimation may be connected with magnetic confinement, or a pressure of external gas [2-4]. Explosive formation of jets in the laboratory is modeled in the experiments with powerful laser beam, and plasma focus [5,6].}

\FullConference{Frontier Research in Astrophysics - III (FRAPWS2018)\\
		28 May - 2 June 2018\\
		Mondello (Palermo), Italy}

\begin{document}

\section{Introduction}

First model of AGN \& quasar, as a supermassive black hole, surrounded by accretion disk, was suggested by D. Lynden-Bell \cite{lb69} in the year 1969.
This model was supported by observations, and
now it is widely accepted that quasars and AGN nuclei contain supermassive black holes (SMBH).
About 10 HMXR (high stellar mass black holes) are found in the Galaxy. They show  behaviour, which, after appropriate scaling,
is similar to the AGN  SMBH, and were called as microquasars \cite {mir92}.

\begin{figure}
\center{\includegraphics[width=5in]{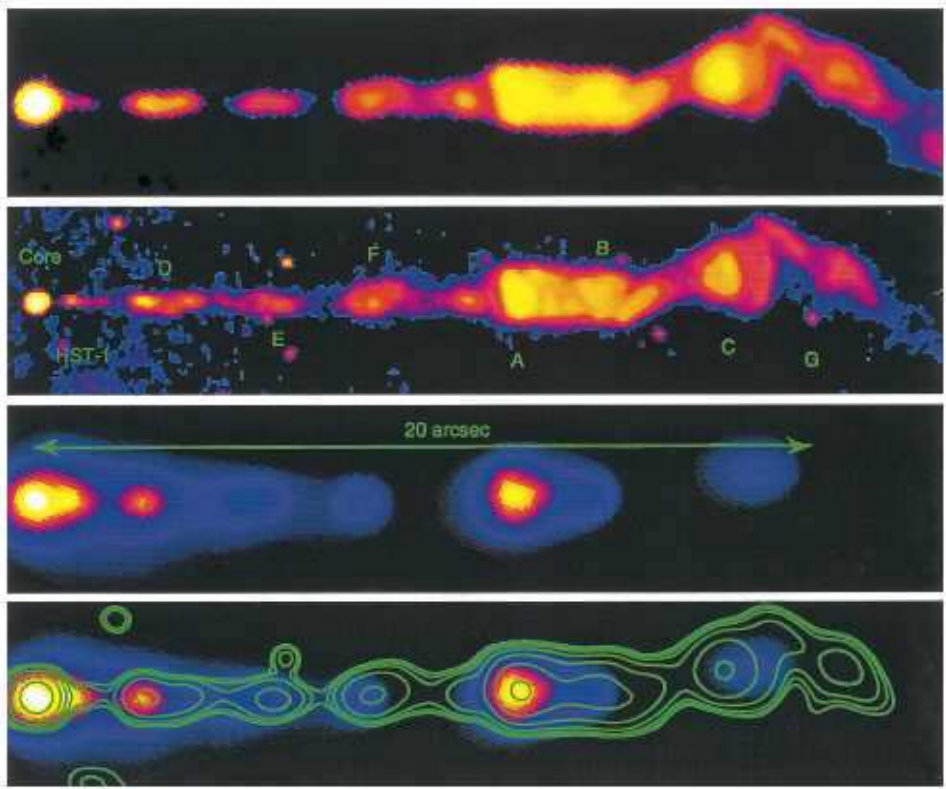}
}
\caption{Images of the jet in M87 in three different bands, rotated to be horizontal, and an overlay of optical contours over the X-ray image. Top: Image at 14.435 GHz using the VLA. The spatial resolution is about 0.$"$2. Second panel: The HST Planetary Camera image in the F814W filter. Third panel: Adaptively smoothed Chandra image of the X-ray emission from the jet of M87 in 0.$"$20 pixels. Fourth panel: Smoothed Chandra image overlaid with contours of a Gaussian-smoothed version of the HST image, designed to match the Chandra point response function. The X-ray and optical images have been registered to each other to about 0.$"$05 using the position of the core. The HST and VLA images are displayed using a logarithmic stretch to bring out faint features, while the X-ray image scaling is linear,
from \cite{mar02}}.
\label{m87col}
\end{figure}

Jets are observed in objects with black holes, where
collimated ejection from accretion disks is expected.
Non-relativistic jets are observed in young stellar-like objects. AGN jets have been studied in many wavebands of the electromagnetic spectrum.
The jet in the Virgo A galaxy M87 was observed in radio (14GHz, VLA) with angular resolution  $\sim 0".2$, in the optics (HST, F814W),
and in soft X-ray band (0.2-8 keV) by Chandra telescope, with angular resolution $\sim 0".2$,
The observations of M87 jet are summarized in \cite{mar02}, and are presented in Fig.\ref{m87col}.

\begin{figure}
\center{\includegraphics[width=4.7in]{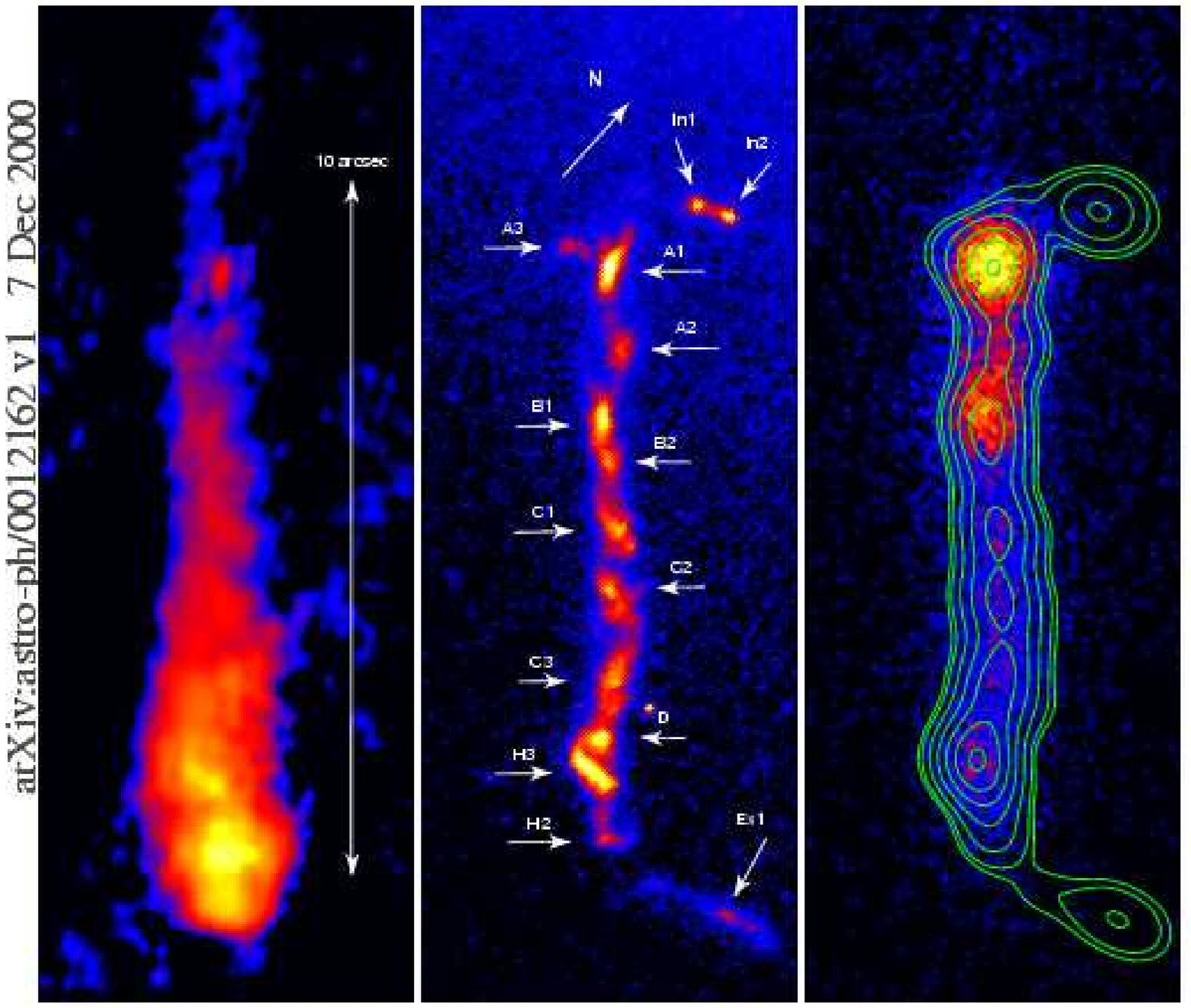}
}
\caption{Images of the jet in 3C 273 in three different bands. Left: Image at 1.647 GHz using the MERLIN array.
Middle: Hubble Space Telescope Planetary Camera image in the F622W filter (centered at 6170 \AA).
Right: Raw Chandra image of the X-ray emission from the jet of 3C 273 in $0".1$ bins overlaid with a version of the HST image
smoothed with a Gaussian profile in order to match the X-ray imaging resolution. The X-ray and optical images have been registered to each other to about $0".05$
using the position of knot A1. The overall shape of the jet is remarkably similar in length and curvature, but the X-ray emission fades to the end of the jet, so
individual C knots are not discernible. The radio emission is much fainter at knot A1 and is displayed with a
logarithmic scaling, from \cite{mar01}
}
\label{3C273}
\end{figure}

The jet in the quasar 3C 273 was observed in radio (MERLIN, 1.647 GHz), in the optics (HST, F622W, centered at 6170A),
and in soft X-ray band (0.2-8 keV) by Chandra telescope, with angular resolution $\sim 0".1$.
The observations of 3C 273 jet are summarized in \cite{mar01}, and are presented in Fig.\ref{3C273}.

\begin{figure}
\center{\includegraphics[width=5in]{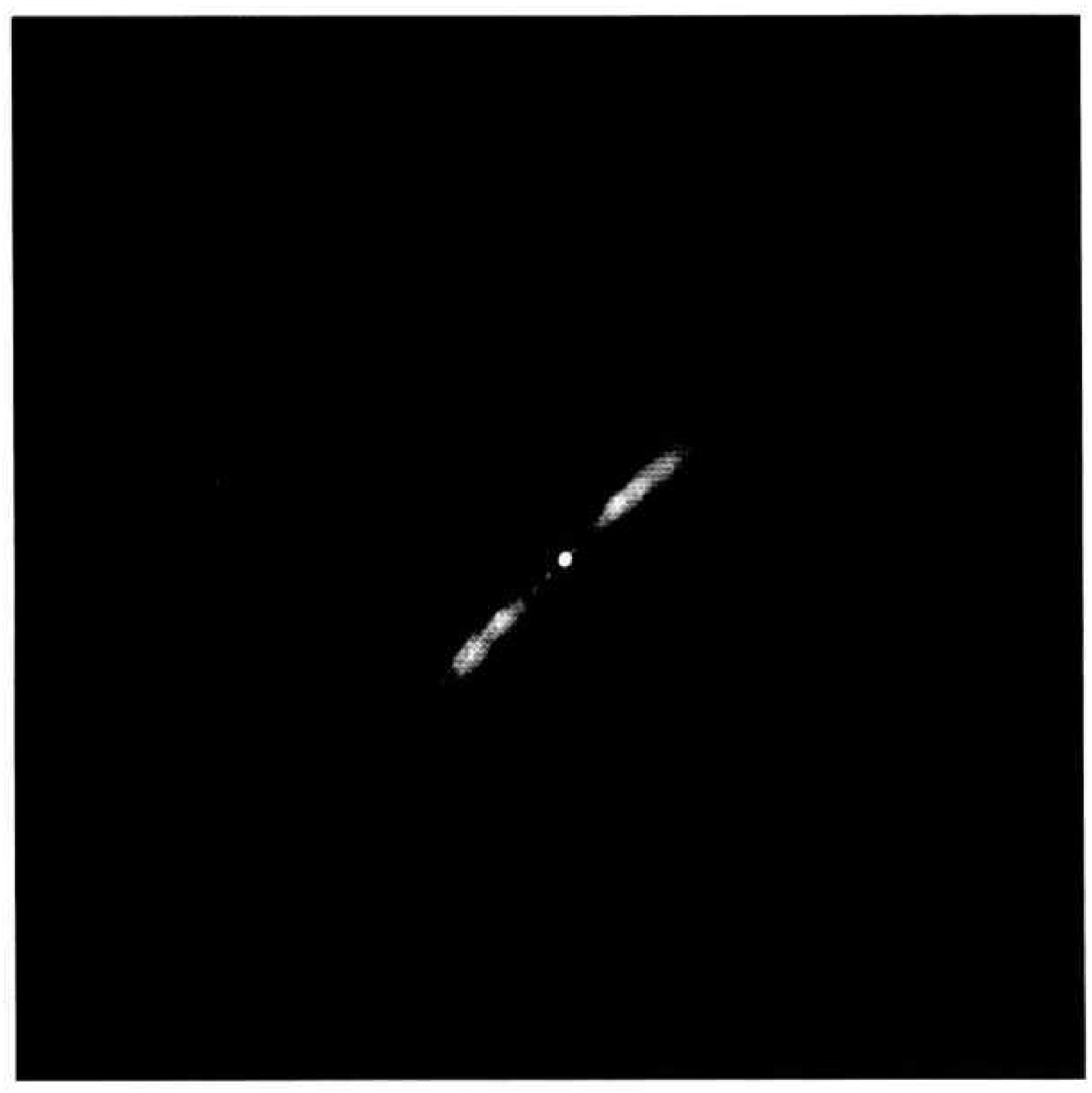}
}
\caption{Jet in the radio galaxy IC 4296 at  20 cm with 3".2 resolution, from
 \cite{kil86}}
\label{jetgal}
\end{figure}

Much longer and fainter
jet  in  radiogalaxy IC 4296 (PKS 1333-33) was observed in radio band  (VLA), at bands between 1.3 and  20 cm, with a best resolution $3".2$ .
The observations of IC 4296 jet are summarized in \cite{kil86}, and are presented in Fig.\ref{jetgal}. Total extent of the jet is about 360 kpc
Radio observations (MERLIN, 5GHz) of a jet ejection in the microquasar GRS 1915+105, have been presented in \cite{fen99}, see Fig.2 in this paper.

\begin{figure}
\center{\includegraphics[width=3in]{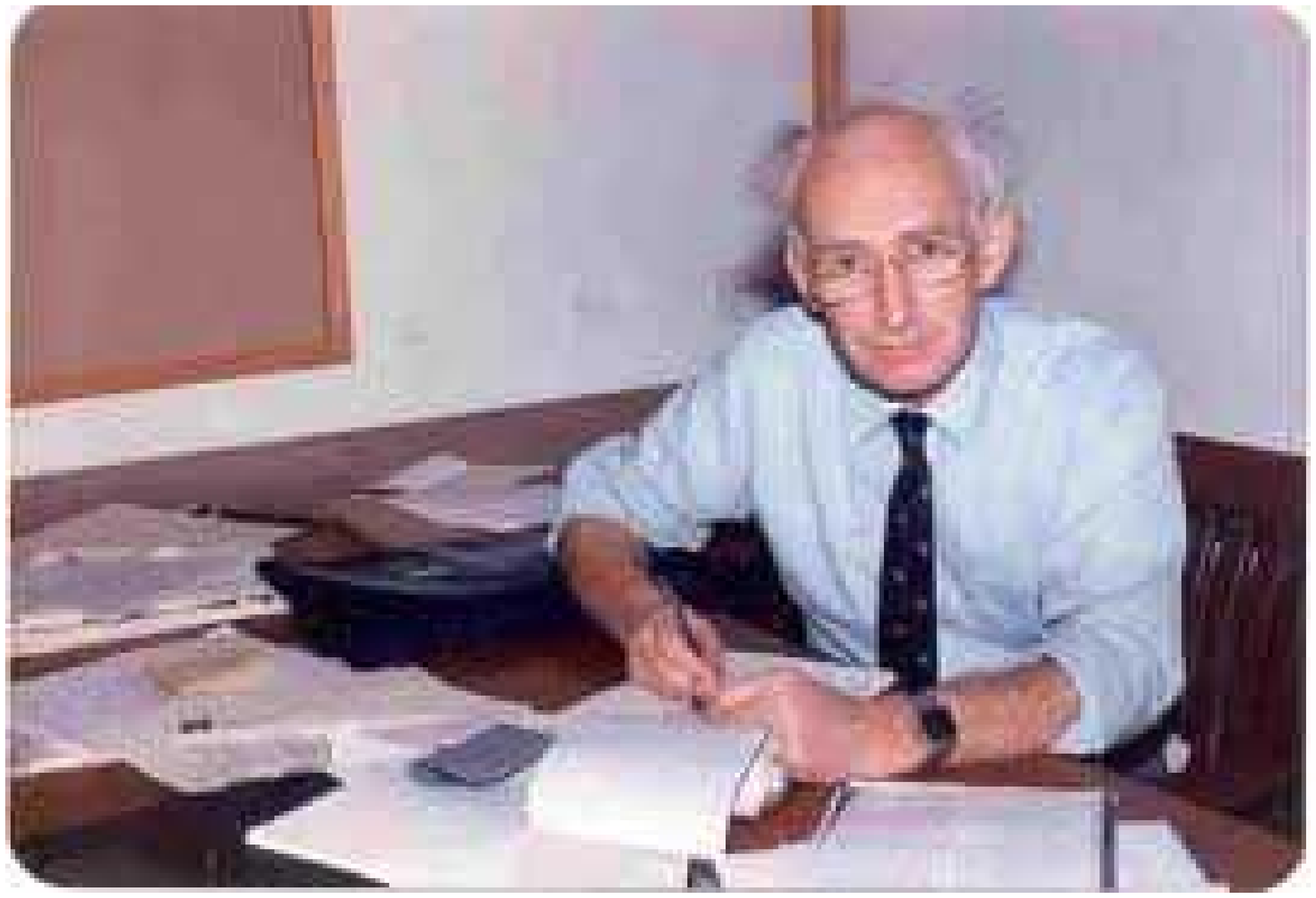}
\includegraphics[width=2.6in]{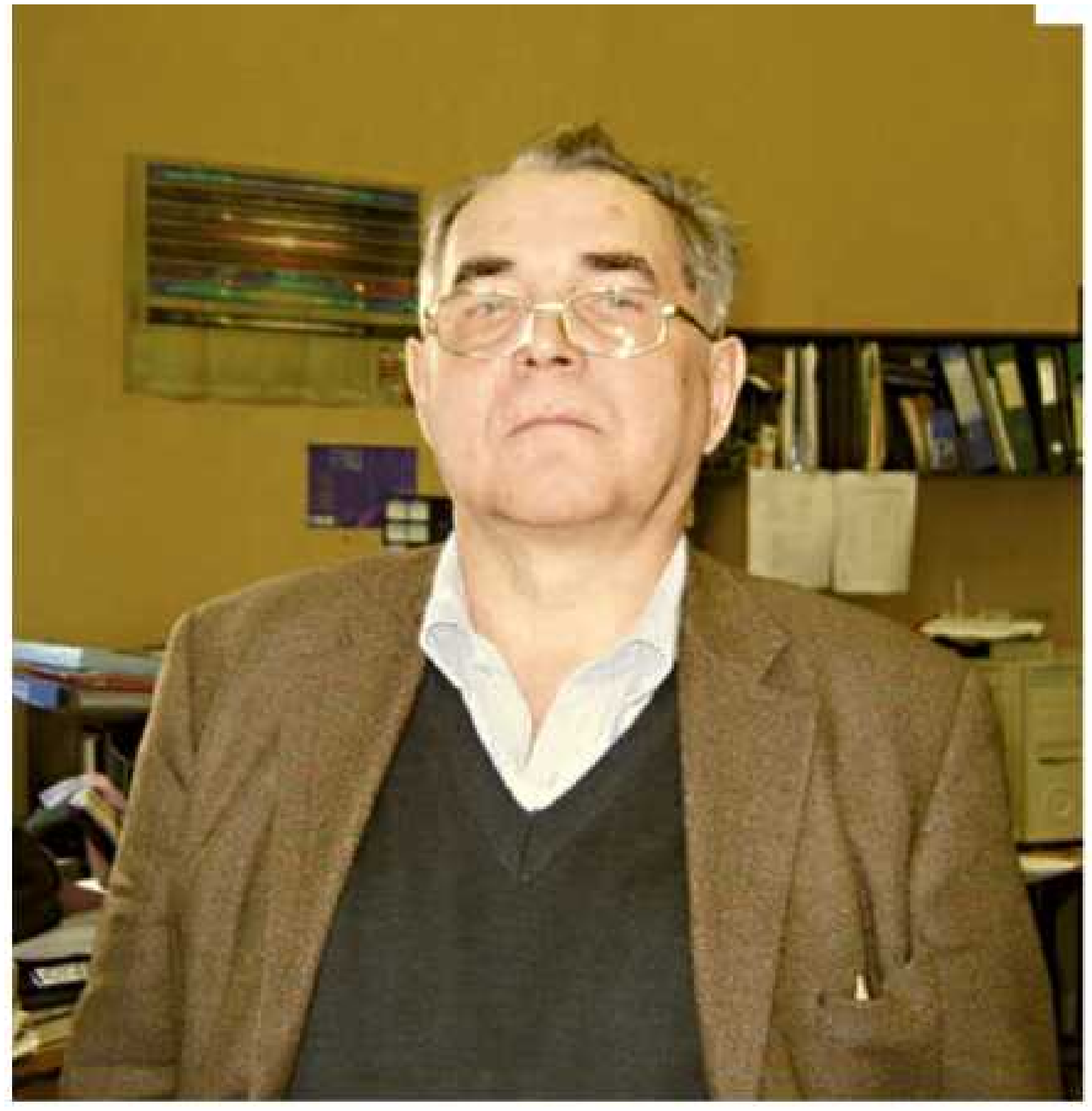}
}
\caption{Donald Lynden-Bell and Nikolay Shakura
}
\label{LBSak}
\end{figure}

\section{Accretion disk models}

\subsection{Large scale magnetic accretion}

\begin{figure}
\center{\includegraphics[width=3.0in]{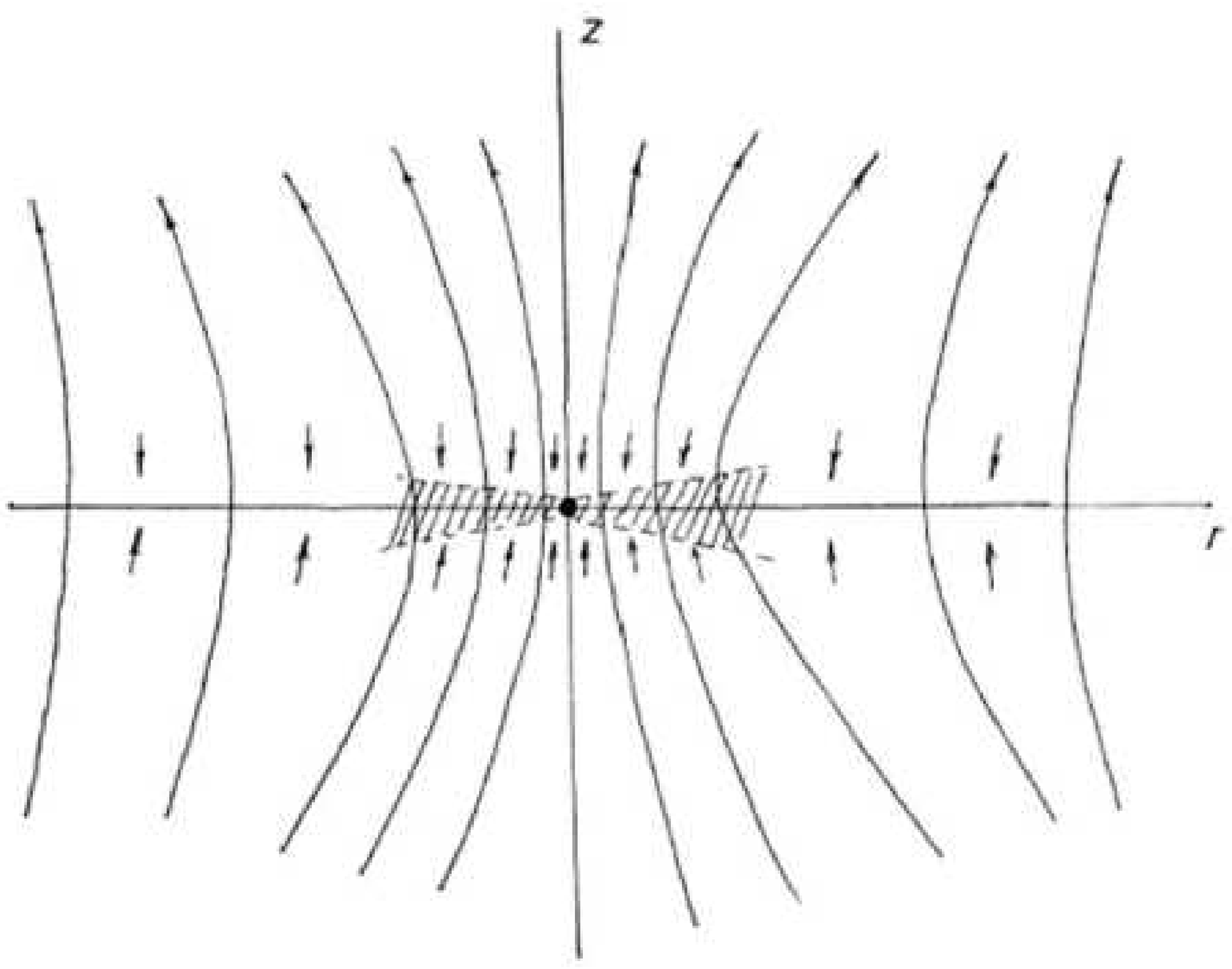}
\includegraphics[width=2.9in]{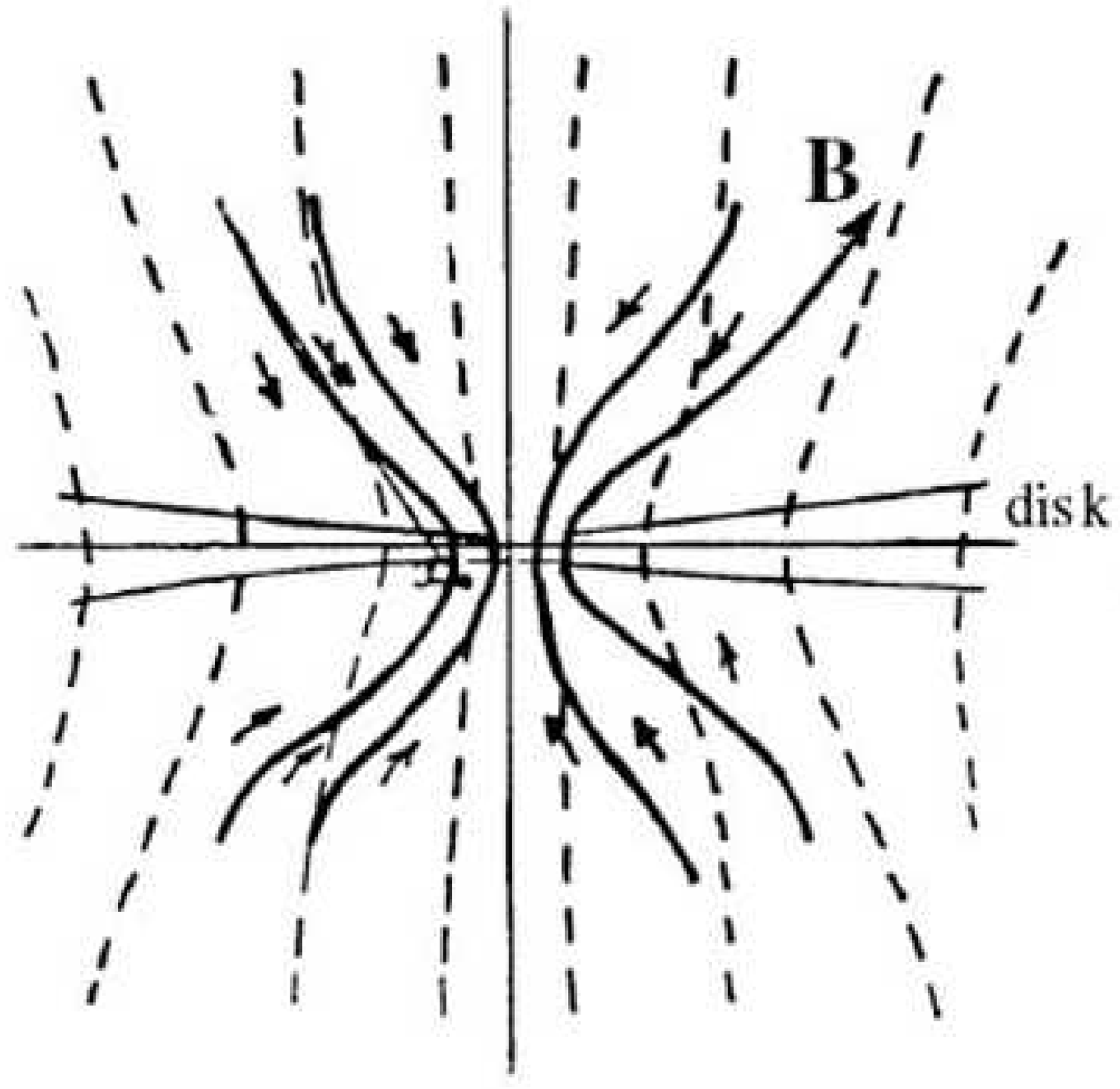}
}
\caption{Sketch of the magnetic field threading an accretion disk in a stationary accretion of non-rotating gas. Increase of the field owing
to magnetic field freezing, in the accreting matter outside the disk, is shown, from
 \cite{bkruz74} (left). Similar picture, with account of the field generated in the non-rotating accretion disk, from \cite{bkruz76}, (right). }
\label{bkruz}
\end{figure}

Accretion disk around BH with large scale  magnetic field (non-rotating disk) had been investigated in \cite{bkruz74,bkruz76}. Self-consistent model of stationary accretion of non-rotating gas, with initially uniform magnetic field, into a black hole (BH) was constructed. At big distances from BH the accretion is almost radial. The magnetic energy, which is small far from BH, in the radial accretion flow is increasing faster than all other types of the energy, and begins to change the flow patterns. Close to the BH the matter flows almost along magnetic field lines, and two flows meat each other in the symmetry plane, perpendicular to the magnetic field direction. Collision of flows leads to development of a turbulence, finite turbulent electrical conductivity, and radial matter flow to BH in the accretion disk, through magnetic field lines. The turbulent electrical conductivity was estimated in \cite{bkruz76} as

\begin{equation}
\label{cond}
\sigma_t = \frac{c^2}{\tilde\alpha 4\pi h \sqrt{P/\rho}},
\end{equation}
analogous to the turbulent $\alpha$ - viscosity in the non-magnetized disks, see next section.
A magnetic field strength in the vicinity of a stellar BH may reach  $10^7 \,-\, 10^{10}$  Gs. At presence of large-scale magnetic field the efficiency of accretion is always large (0.3-0.5) of the rest mass energy flux. Formation of nonrotating disk around BH, supported by magnetic field strength, and self-consistent picture of accretion with account of magnetic field created by induced toroidal electrical currents in the accretion disk are presented in Fig. \ref{bkruz}.

\subsection{Standard accretion  disk model}

Algebraic relation for construction of the thin accretion disk model were used in \cite{lb69}. Another presentation, based on so called "alpha disk" model, suggested in the paper \cite{shak72}, was more attractive, and was used in different types of accretion disks.
The small thickness of the disk in
comparison with its radius $h \ll r$
indicate to small importance
of the pressure gradient
$\nabla P$ in comparison with
gravity and inertia forces.
That leads to a simple
radial equilibrium equation
denoting the balance between the last two
forces occurring when the angular
velocity of the disk $\Omega$ is equal to
the Keplerian one $\Omega_K$,

\begin{equation}
\label{ref1.1}
\Omega=\Omega_K=\left(\frac{GM}{r^3}\right)^{1/2}.
\end{equation}
For a thin disk the differential
equation for a vertical equilibrium is substituted by an
algebraic one, determining the
half-thickness of the disk in the form

\begin{equation}
\label{ref1.3}
h \approx \frac{1}{\Omega_K}
\left(2\frac{P}{\rho}\right)^{1/2}.
\end{equation}
The balance of angular momentum,
related to the $\varphi$ component of the
Euler equation has an integral
in a stationary case written as

\begin{equation}
\label{ref1.4}
\dot M(j-j_{in})=-2\pi r^2\,2ht_{r\varphi},\quad t_{r\varphi}=
\eta r\frac{d\Omega}{dr}.
\end{equation}
Here $j=\upsilon_{\varphi}r=\Omega r^2$ is
the specific angular momentum,
$t_{r\varphi}$ is a component of the
viscous stress tensor, $\dot M>0$ is a mass
flux per unit time into a black hole,
$j_{in}$ is an integration constant  equal to the specific angular
momentum of matter
falling into a black hole.

\begin{equation}
\label{ref1.5}
j_{in}=\Omega_K r_{in}^2,
\end{equation}
In the $\alpha$ disk model ($\upsilon_t=\alpha\upsilon_s $), $\upsilon_t$ is an average turbulent velocity, $\upsilon_s=\sqrt{P/\rho}$ is a sound speed, the component $t_{r\varphi}$ of the stress tensor is
taken \cite{shak72} proportionally to the isotropic pressure, so that

\begin{equation}
\label{ref1.8}
t_{r\varphi}=\rho \upsilon_t h r \frac{d\Omega}{dr}
\approx \rho \upsilon_t \upsilon_s =-\alpha
\rho \upsilon_s^2= -\alpha P, \quad \eta=\rho\upsilon_t h.
\end{equation}

\begin{figure}
\center{\includegraphics[width=2.6in]{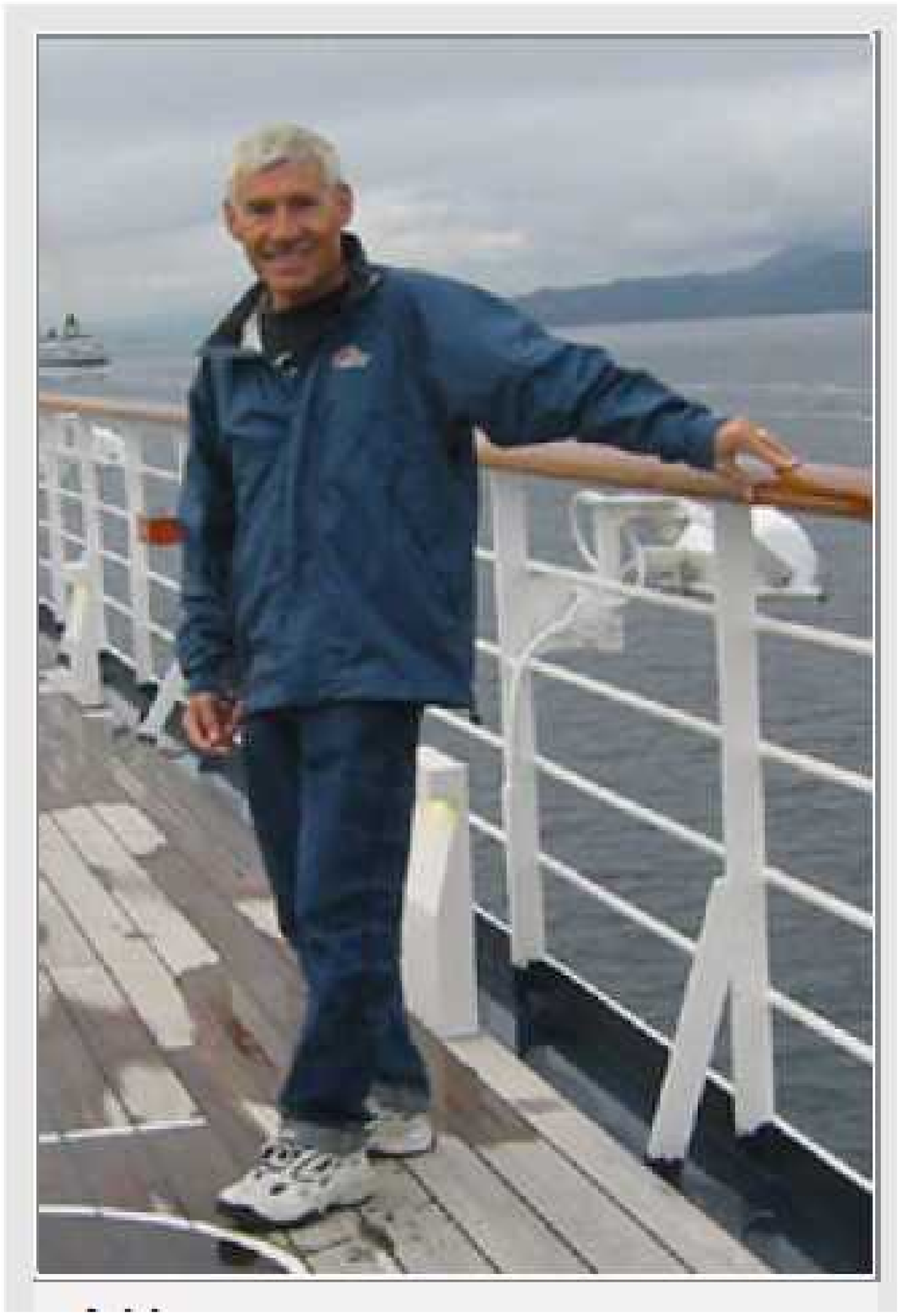}
\includegraphics[width=3in]{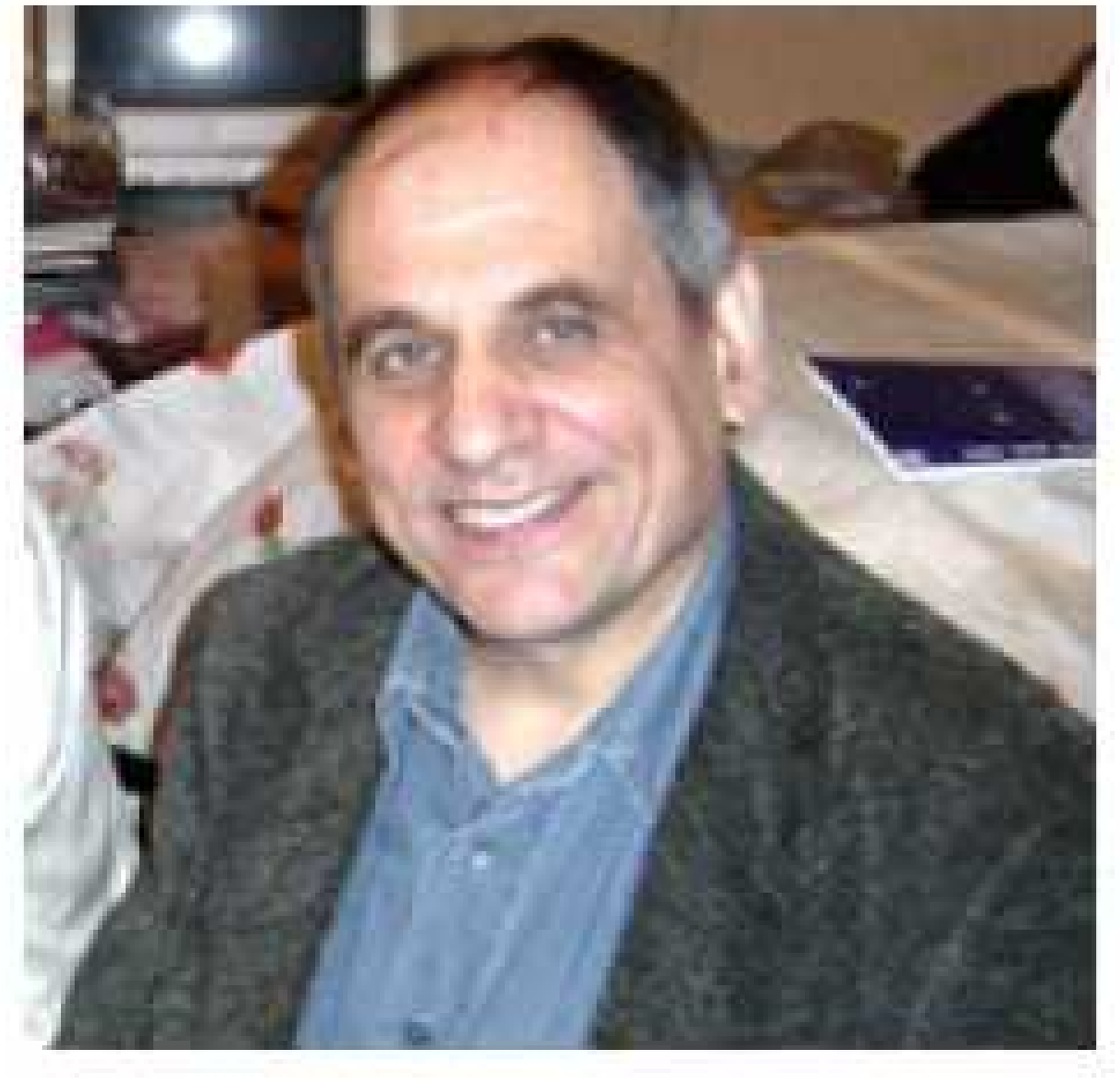}
}
\caption{Alexandr Ruzmaikin and Sergey Blinnikov
}
\label{RB}
\end{figure}

\subsection{Convection and hot corona}

\begin{figure}
\center{\includegraphics[width=0.9\linewidth]{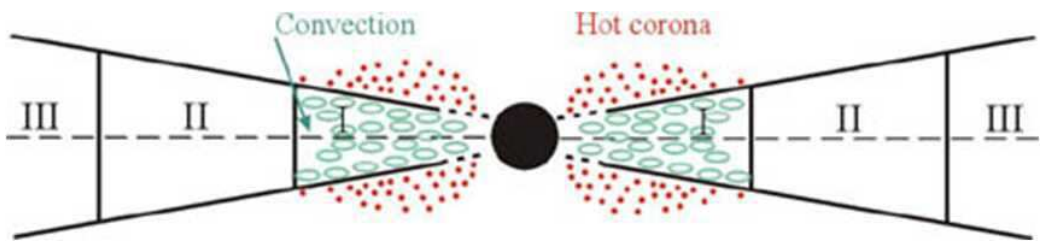}}
\caption{Sketch of picture of a disk accretion on to a black hole
at sub-critical luminosity.
\indent\indent I - radiation dominated region, electron scattering.
\indent\indent II - gas-dominated region, electron scattering.
\indent\indent III - gas-dominated region, Krammers opacity; convective region,
 and hot corona are indicated, from \cite{bk85}.
}
\label{adisk}
\end{figure}

Self-consistent structure of the optically thick accretion disk around a black hole has three regions, depending on the origin of pressure and opacity \cite{ss73}, see Fig.\ref{adisk}.
It was shown in \cite{bkb76}, that inner, radiation dominated regions of the accretion disk are convectively unstable, and, therefore, produce a hot corona with electron temperature about $10^8\,-\,10^9$ K. The model of accretion disk with a hot corona was used for explanation of
properties of the X ray source Cygnus X-1, and transition between different states in this source \cite{bkb76}.

\begin{figure}
\center{\includegraphics[width=0.5\linewidth]{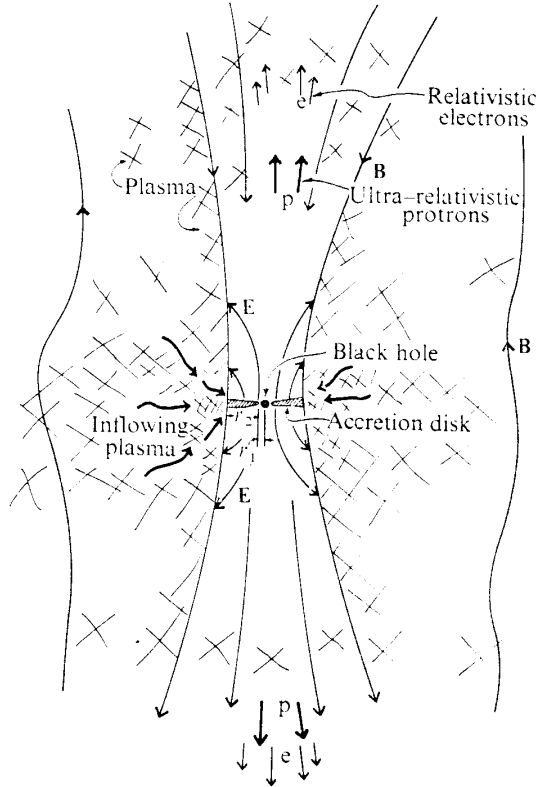}}
\caption[h]{Sketch of the electromagnetic outflows from
the two sides of the disk owing to
the Faraday unipolar dynamo action
of a rotating magnetized disk, from \cite{lov76}.
}
\label{fjet2}
\end{figure}

  A presence of a large scale magnetic field in the inner regions of a keplerian accretion disk create mechanism of particle ejection and jet formation \cite{bkb76}.
Mechanism for producing fast particles  is analogous to the pulsar process. If magnetized matter with low angular momentum falls into the black hole (in addition to the disk accretion), a strong poloidal magnetic field will arise. By analogy to pulsars, rotation will generate an electric field of strength $E \approx −(\upsilon/c)B$ in which electrons are accelerated to energies $\approx R(\upsilon/c)B\, e \approx 3 \cdot 10^4 [B/(10^7 {\rm Gauss})]$ Mev where $\upsilon/c \approx 0.1$ and $R \approx 10^7$ cm is the characteristic scale. In a field $B \approx 10^7$ Gauss, such electrons will generate synchrotron radiation with energies up to $\approx  10^5$ keV. It would be possible here for $e^+e^−$ pairs to be formed and to participate in the synchrotron radiation. The flow of high-energy particles along the magnetic field should be visible as a highly collimated flow - jet. Similar model of a jet formation, with account of dynamo processes was considered in the paper \cite{lov76}, see Fig.\ref{fjet2}.

\begin{figure}
\center{\includegraphics[width=2.6in]{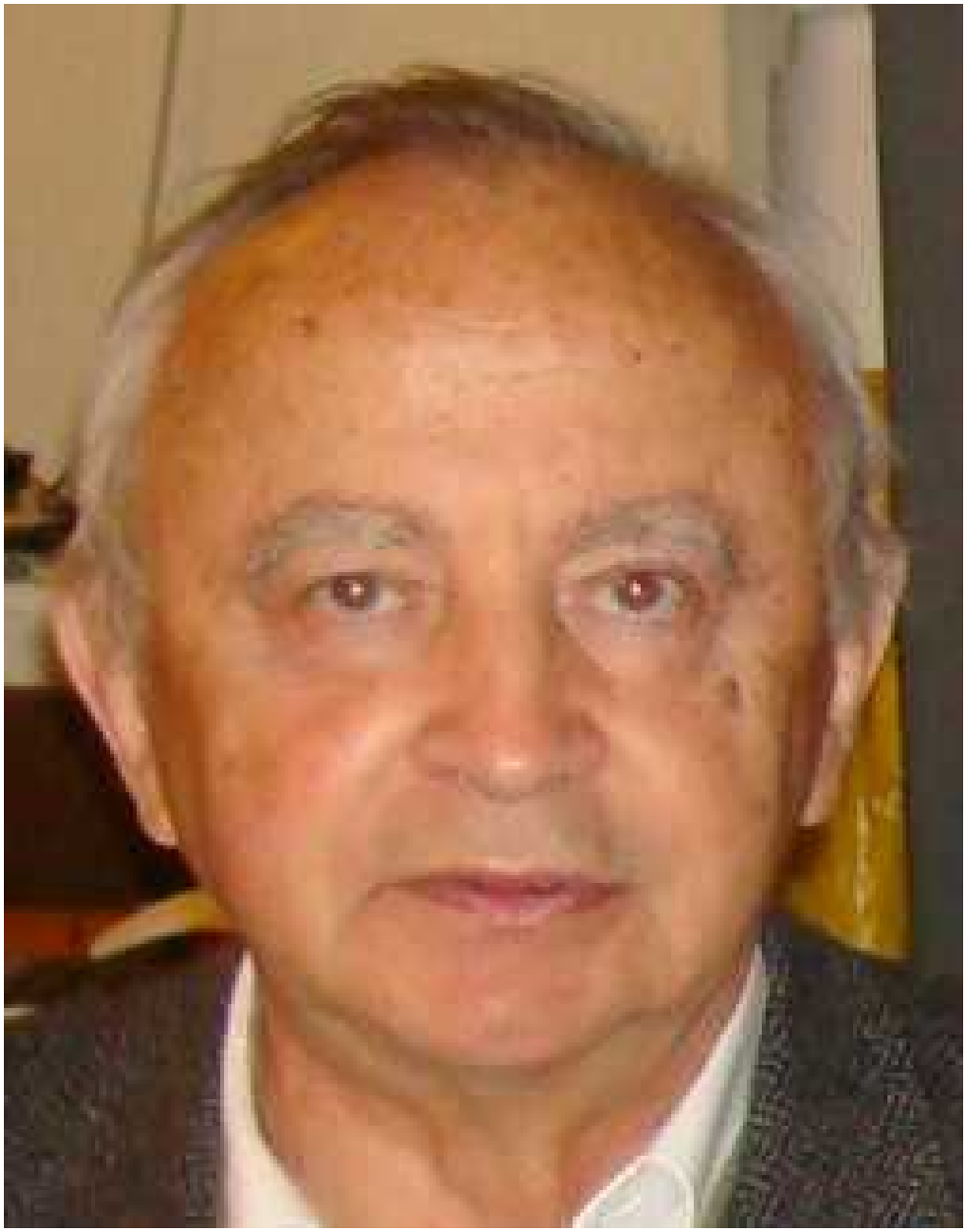}
\includegraphics[width=3in]{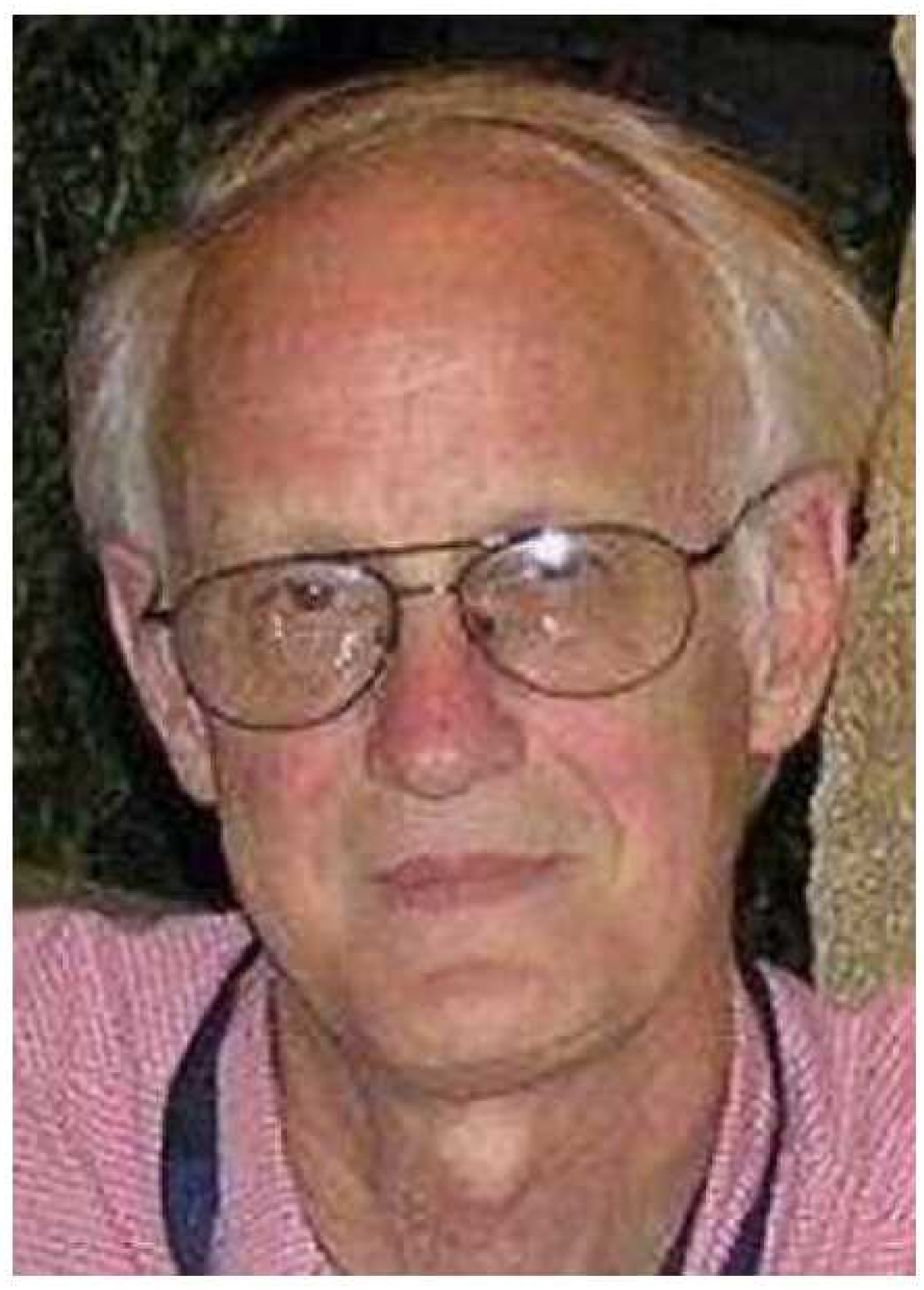}
}
\caption{Igor Novikov and Richard Lovelace
}
\label{NL}
\end{figure}

\section{Solutions with advection}

\begin{figure}
\begin{center}{\includegraphics[width=2.9in]{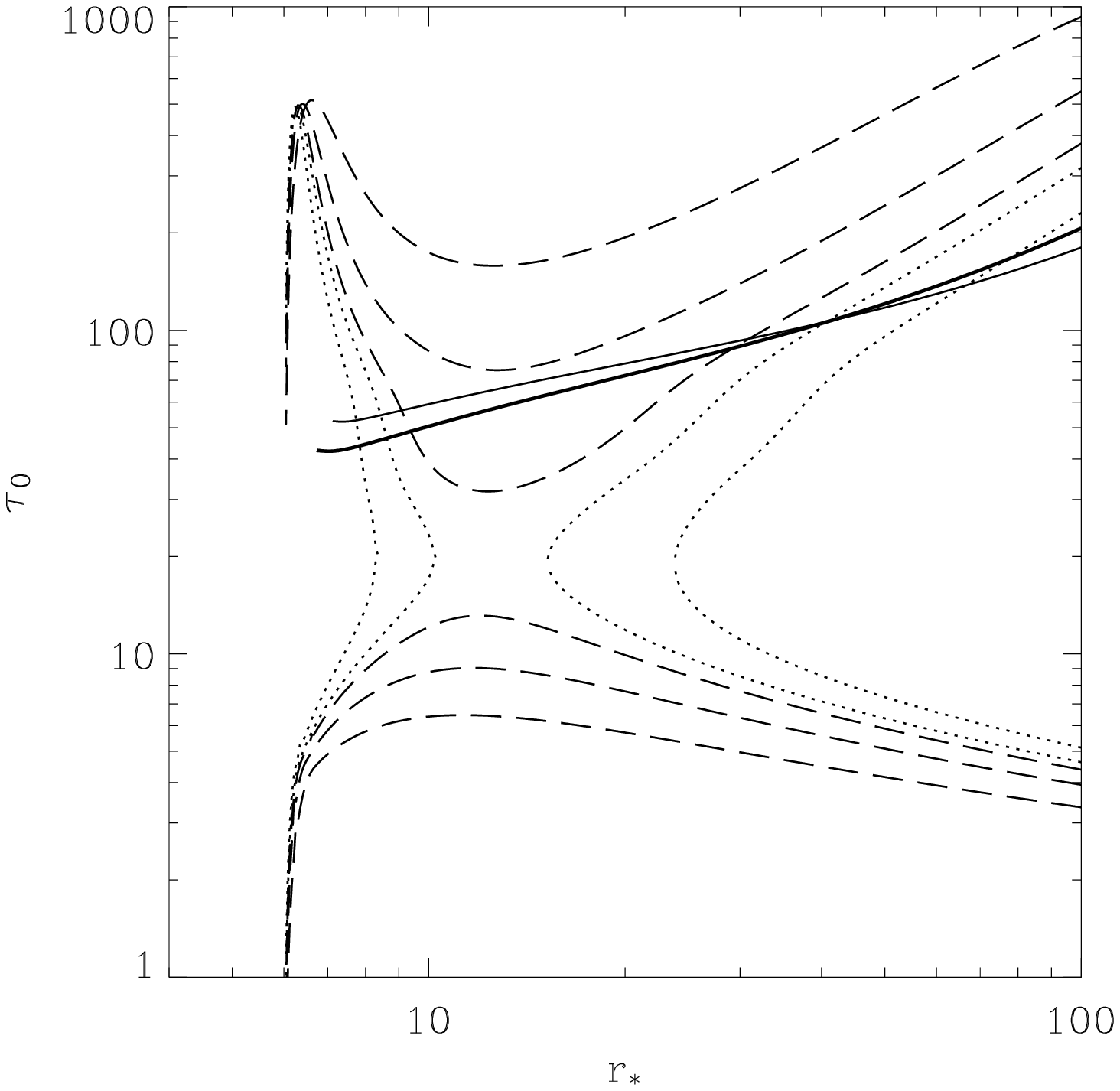}
\includegraphics[width=2.9in]{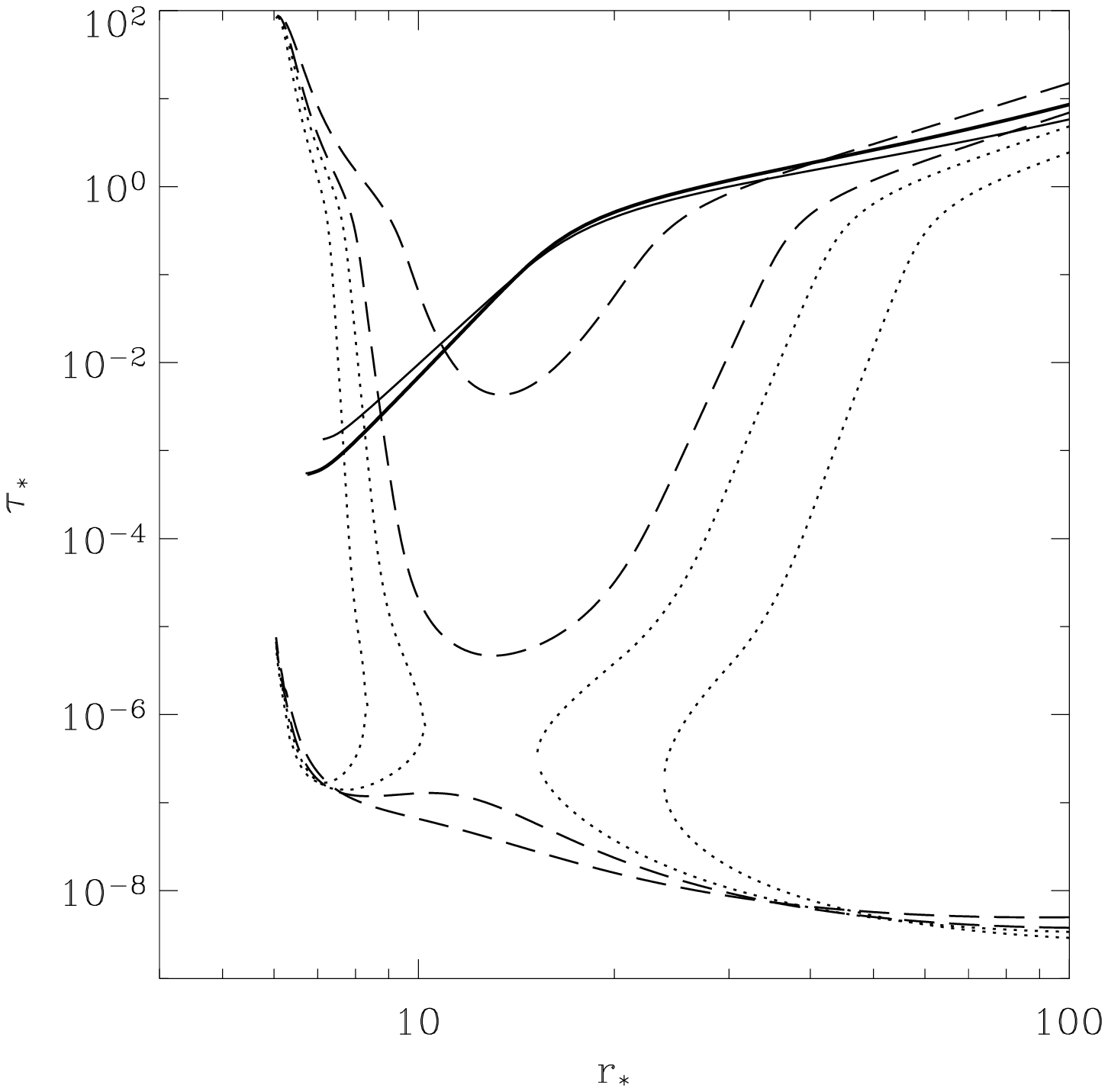}}
\end{center}
\caption{The dependence of
the Thompson scattering depth (left), and the effective
optical depth (right) on the radius , and
for the models with $\alpha=0.5$ and
$M_{BH}=10M_{\odot}$.
Dashed lines correspond to the solutions without advection and
$\dot{m}<\dot{m}_{cr}=36$.
Dotted lines correspond to the non-physical solutions without
advection for $\dot{m}=\dot{m}_{cr}=36$ and $\dot{m}=50$ (from the
center to the edge of the picture respectively). Solid lines
correspond to the solutions with advection and the mass accretion
rate higher than the critical one. Thick solid line corresponds to
$\dot{m}=36.0$, and the thin solid line to $\dot{m}=50.0$,
$\dot{m}=\dot M/\dot M_{\rm Edd}$, where $\dot M_{\rm Edd}=L_{\rm
Edd}/c^2,\,\,\, L_{\rm Edd}=4\pi c G  M m_p/\sigma_T$;
$r_*\equiv x$, from \cite{abkin1}.
\label{fig1}}
\end{figure}

\begin{figure}
\center{\includegraphics[width=4in]{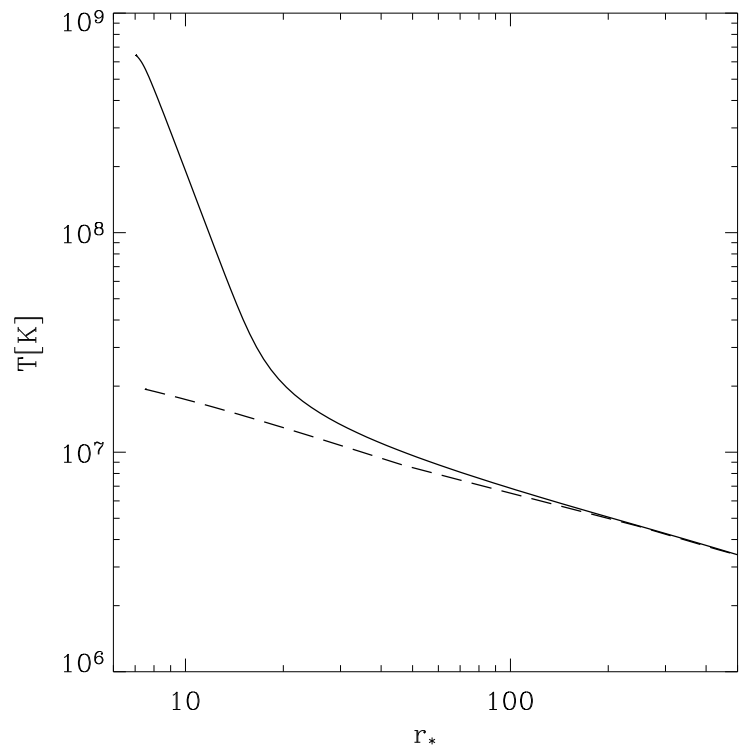}}
\caption{The dependence of the temperature on the
radius for the same models as in Fig.\ref{fig1}; $r_*\equiv x$, from \cite{abkin1}.
\label{fig6}}
\end{figure}

Standard (local) accretion theory is not correct at luminosity, approaching Eddington limit, and in the vicinity of a last stable orbit around BH. Advection of energy along the accretion disk was taken into account in \cite{pbk}. In very luminous accretion disks only advective models, qualitatively different from standard ones, give proper results. These models are characterized by decrease of a vertical optical depth with decreasing of a radius, so that
accretion disk is optically thick at larger, and optically thin at smaller radiuses, with a gradual transition between these regions \cite{abkin,kbk}.
Set of equations for "$\alpha$ P" viscosity prescription

\begin{equation}
\label{ad1}
t_{r\varphi}= \alpha P,
\end{equation}
where $\alpha$ is a parameter, $\alpha \leq 1$, with advection, had been solved in \cite{abkin}.
Radiative cooling term $Q^{-}$, and equation of  state, describing smooth  optically thick-thin transition were take from \cite{abkbn} as

\begin{equation}
\label{ad2}
Q^{-}={2 a T^4 c
\over 3 \tau_{0}}\left(1+\frac{4}{3\tau_{0}}+
\frac{2}{3\tau_{*}^2}\right)^{-1},
\end{equation}

\begin{equation}
\label{ad3}
P=\rho{\cal R}T\,+\,{a T^4
\over 3}\left(1+\frac{4}{3\tau_{0}}\right)
\left(1+\frac{4}{3\tau_{0}}\textsl{+}\frac{2}{3\tau_{*}^2}\right)^{-1}.
\end{equation}
Here $\tau_{0}=\kappa \rho h$ is a total Thomson scattering
depth of the disk and

\begin{equation}
\label{ad4}
\tau_{*}=\left(\tau_{0}\tau_{\alpha}\right)^{1/2}
\end{equation}
is the effective optical depth valid
for the case $\tau_{0} \gg \tau_{\alpha}$, which takes place in a
region with intermediate optical depths, $\tau_{\alpha}$ is the
optical depth with respect to bremsstrahlung absorption,

\begin{equation}
\label{ad5}
\tau_{\alpha}\simeq 5.2\cdot 10^{21}{\rho^{2}T^{1/2}h\over acT^{4}} ~.
\end{equation}
 A numerical solution of the set of the following non-dimensional equations was obtained in \cite{abkin}. Gravitational potential $\varphi_g$ of Paczynski-Witta \cite{pw} was used, accounting for some effects of general relativity:

 \begin{equation}
\label{ad6a}
\varphi_g=\frac{GM}{r-2r_g},
\end{equation}

\begin{equation}
\label{ad6}
 r{\upsilon'\over \upsilon}={N\over D} ~,
\end{equation}

\begin{equation}
\label{ad7}
r{c_s'\over c_s}=1+\frac{c^2\,x^2}{{c}_s^2}\left(\tilde{\Omega}^2-
\frac{1}{x(x-2)^2}\right)+\frac{3x-2}{2(x-2)}-
\left(\frac{{\upsilon}^2}{{c}_s^2}-1\right){N\over D} ~~,
\end{equation}
where the notations $N$ and $D$ denote algebraical
expressions, depending only on $r$, $\upsilon$, $c_s$, and $\ell_{\rm
in}$.

\begin{equation}
\label{ad8}
\Omega={\ell_{in}\over r^2}+\alpha{c_s^2\over \upsilon r} ~.
\end{equation}

\begin{equation}
\label{ad9}
(1-\beta)P={aT^4\over 3}\left(1+\frac{4}{3\tau_{0}}\right)
\left(1+\frac{4}{3\tau_{0}}\textsl{+}\frac{2}{3\tau_{*}^2}\right)^{-1}.
\end{equation}
\begin{equation}
\label{ad10}
x=\frac{r}{r_g}, \quad r_g=\frac{GM}{c^2},\quad
\widetilde{\Omega}=\frac{\Omega r_{g}}{c}, \quad
\beta=\frac{P_g}{P}=\frac{{\cal R}T}{c_s^2}, \quad
c_s^2=\frac{P}{\rho}, \quad
T=\beta\frac{c_s^2}{\cal R}
\end{equation}
Here $\ell_{in}$ is a specific angular momentum on the last stable orbit.
Solution of equations (\ref{ad6a})-(\ref{ad7}), with account of (\ref{ad8})-(\ref{ad10}), for $\tau_{0}(r_*)$, $\tau_{*}(r_*)$, $T(r_*)$ are given in Figs.\ref{fig1},\ref{fig6}, from \cite{abkin1}.

\noindent In a rotating BH, with the Kerr metric, the temperature in the optically
thin region exceeds 500 keV, when an intensive $e^+e^-$  pair creation takes place \cite{bk11a}.

\begin{figure}
\center{\includegraphics[width=2.6in]{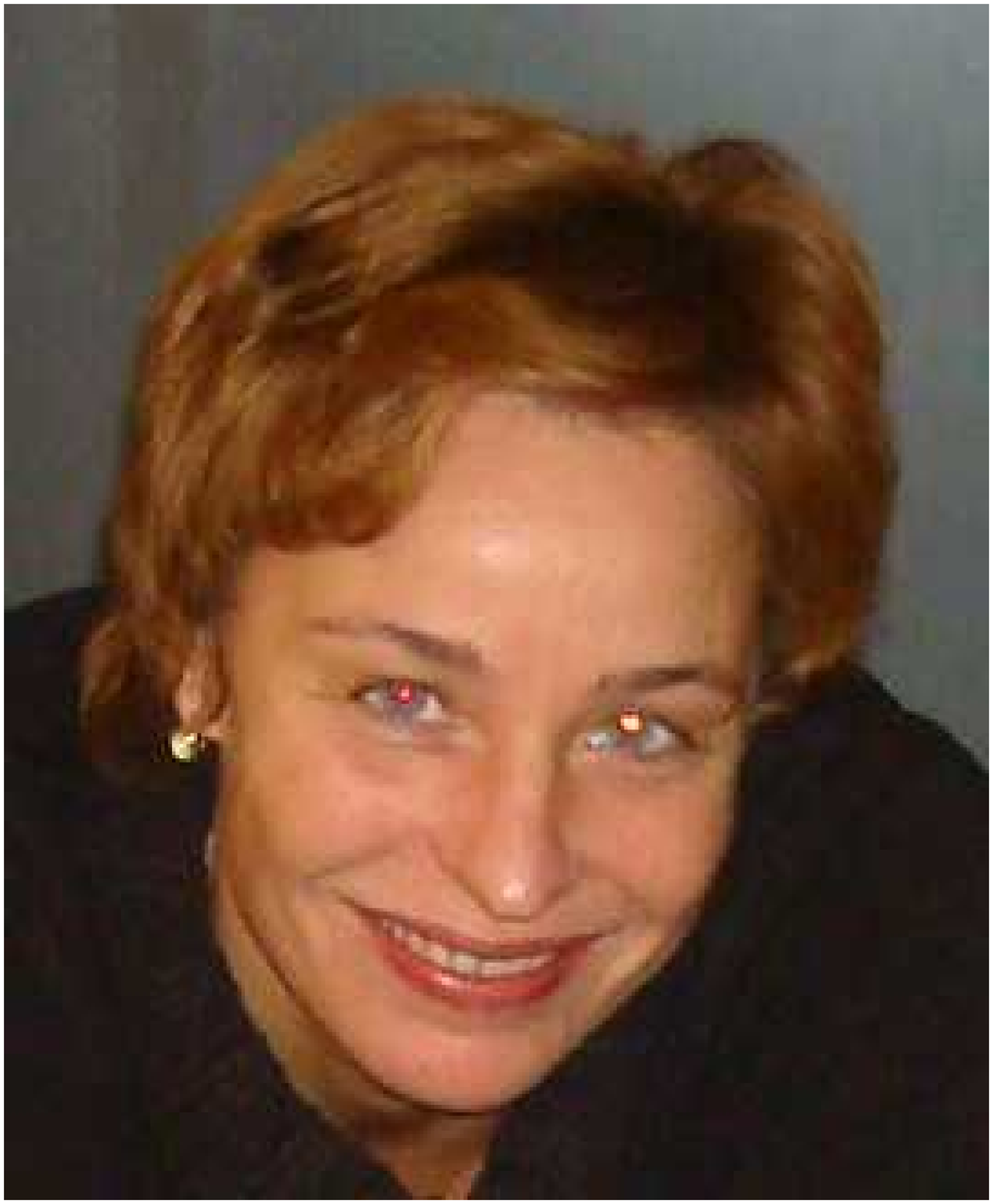}
\includegraphics[width=3in]{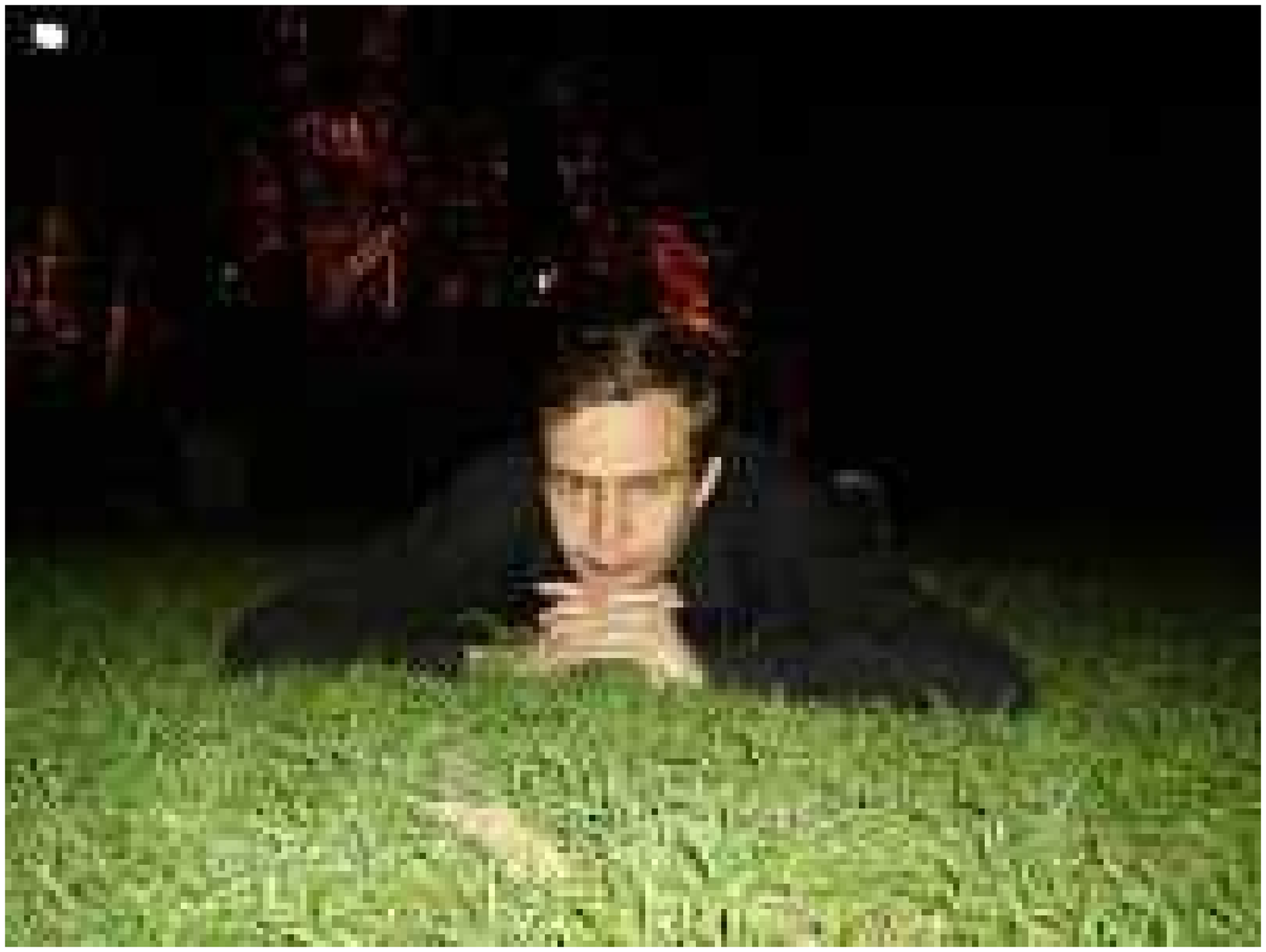}
}
\caption{Julia Artemova and Alexandr Klepnev
}
\label{AK}
\end{figure}

\section{Magnetic jet collimation}

\begin{figure}
\center{\includegraphics[width=2.9in]{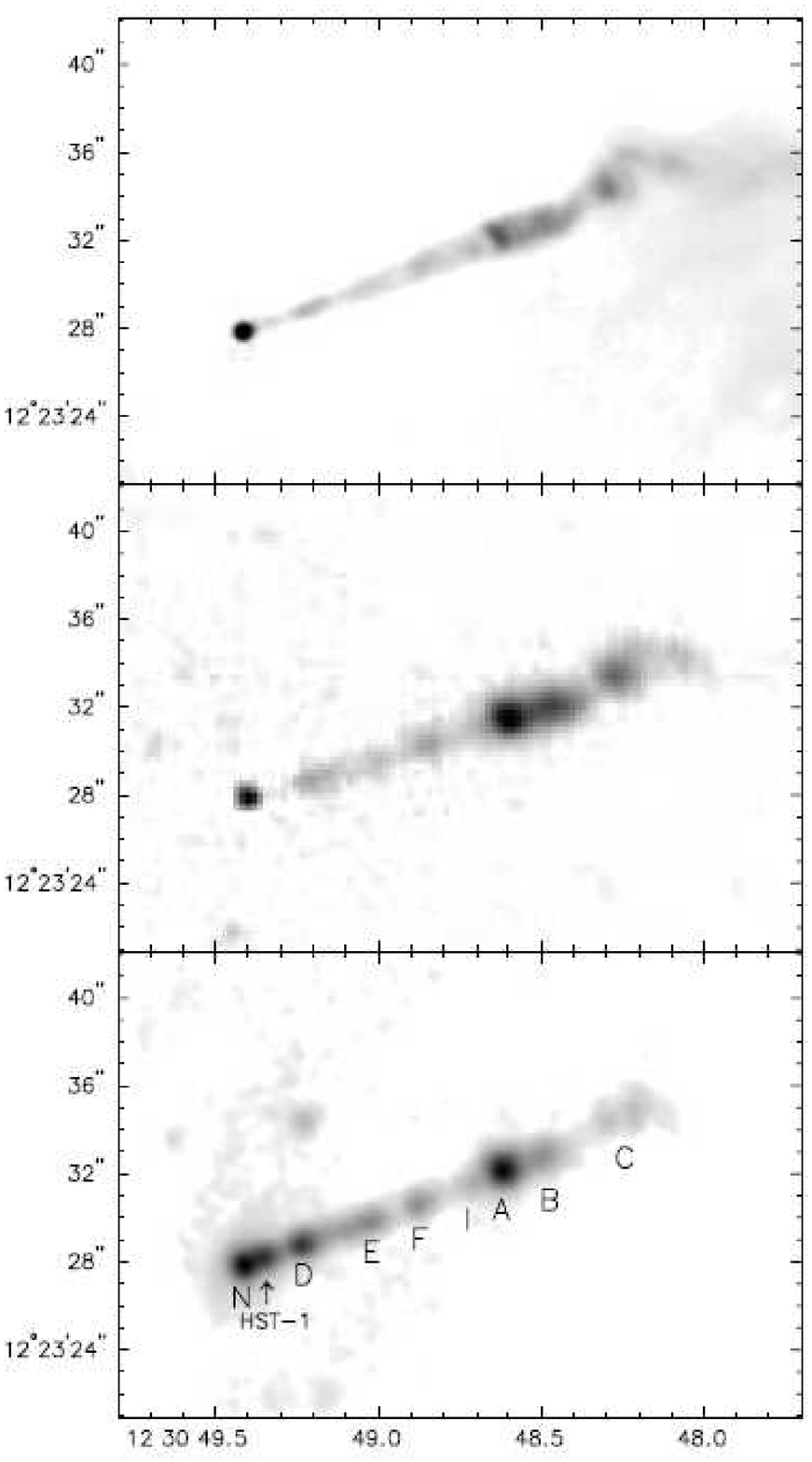}
\includegraphics[width=1.5in]{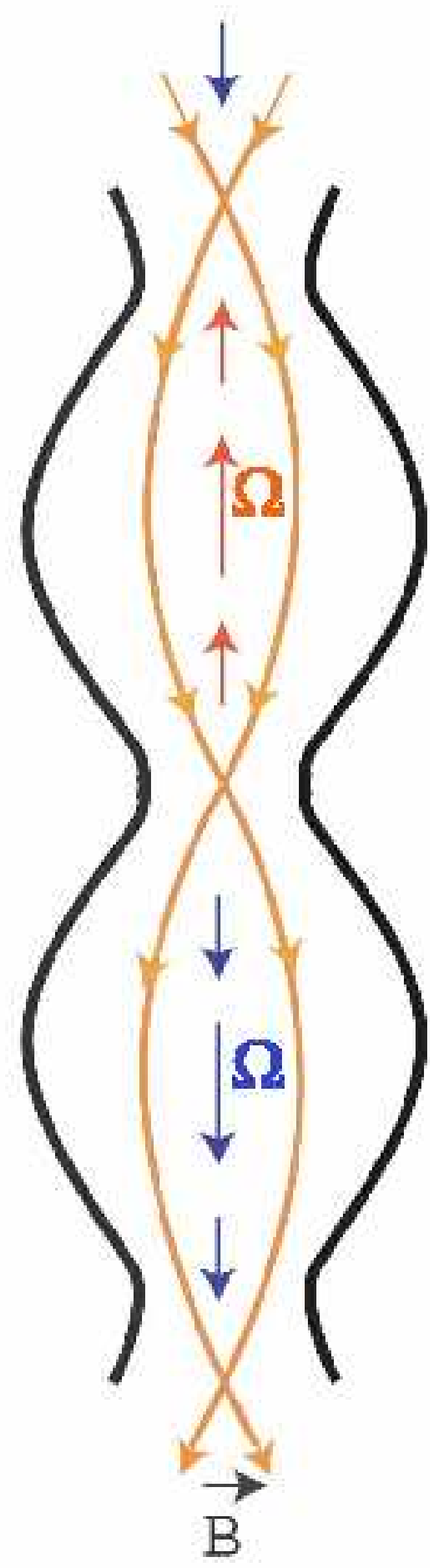}
}
\caption{Gray-scale representations of a 6 cm radio (top; resolution 0$"$4);
 an optical V band (middle, resolution
0$"$7);  and the Chandra
X-ray (bottom, resolution 0$"$7; 0.1 - 10 keV band) image.
The labels in the lower panel refer to the knots vertically above the label. N
is the nucleus, from \cite{wy02} (left).
Scheme of magnetic collimation due to torsional oscillations, from \cite{bk07} (right).}
\label{collin}
\end{figure}

Observations of extragalactic jets in different objects distinctly show existence of bright knots along a whole jet in different wavelengths, Fig.\ref{collin}, see also Figs.\ref{m87col},\ref{3C273}.

\begin{figure}
\center{\includegraphics[width=2.9in]{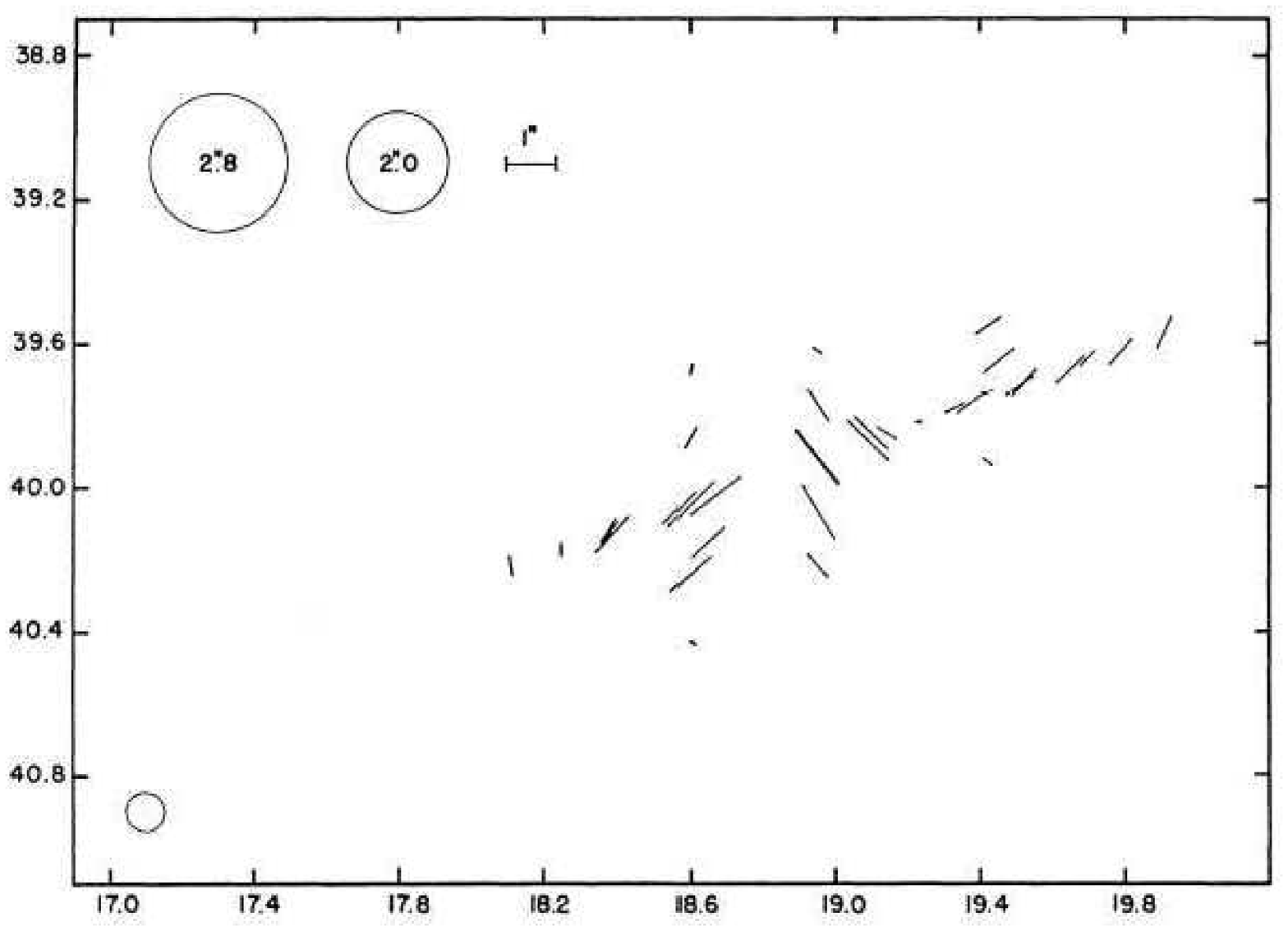}
\includegraphics[width=3in,angle=00]{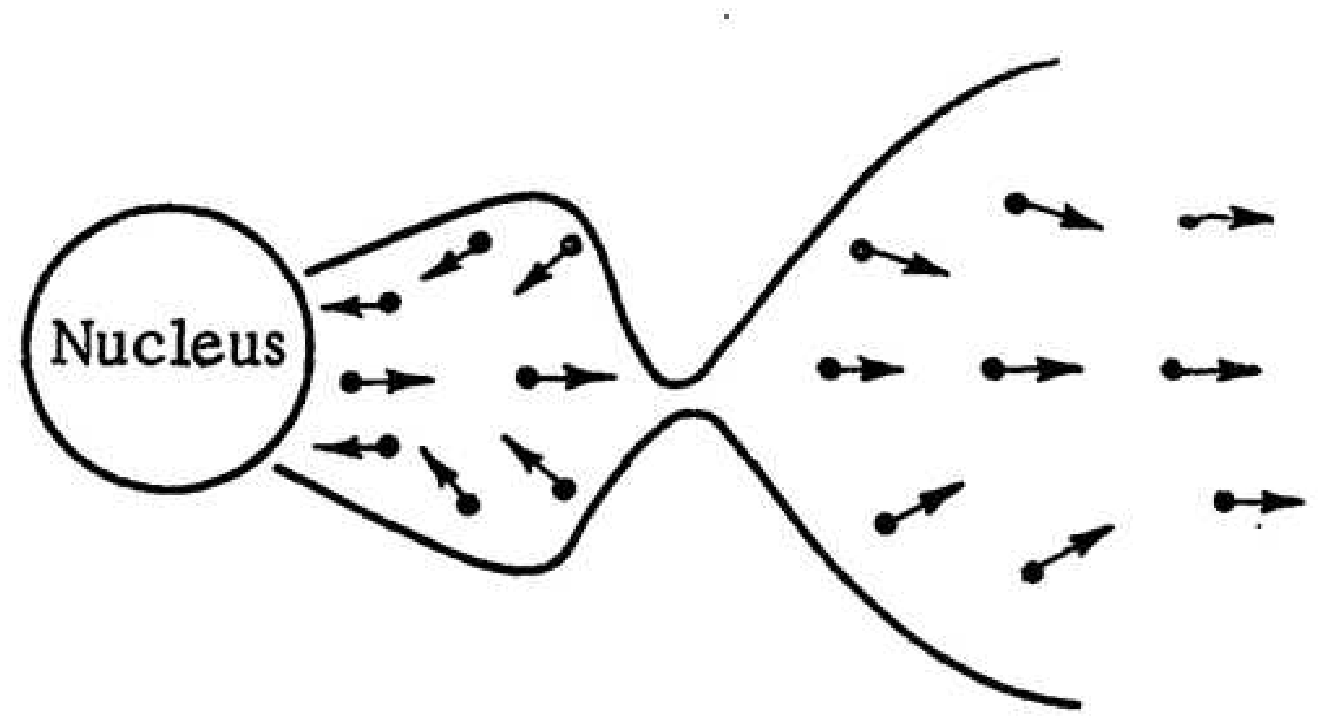}
}
\caption{ Photoelectric polarization observations of the jet in M87.
The polarization angles in neighboring blobs are orthogonally related. The relative sizes of the diaphragms used are shown in the upper left of the figure. The position of the nucleus of M87 is shown by a small open circle in the lower left, from \cite{hil} (left).
Magnetically collimated jet by inductance-capacitance oscillations in the blob, from \cite{bkf} (right).
}
\label{polar}
\end{figure}

It was shown by optical photoelectric polarization observations of the jet in M87 \cite{hil},
that polarization angles in neighboring blobs are orthogonally related, see Fig.\ref{polar}.

This behavior was interpreted in the model of magnetic collimation \cite{bkf}, where initial charge separation in the neighboring blobs leads to oscillating electrical current, as in a capacitance-inductance system. This current produces azimuthal
magnetic field, preventing jet expansion and disappearance. The period of oscillations should be $P_{osc} \sim D/c$, where $D$ is a size of the blob in the jet, see Fig.\ref{polar}.

Magnetic collimation, connected with torsional oscillations of a cylinder with
elongated magnetic field, was considered in \cite{bk07}. Instead of initial blobs with charge separation, there is
 a cylinder with a periodically distributed initial rotation
around the cylinder axis. The stabilizing azimuthal magnetic field is created here by
torsional oscillations, where charge separation is not necessary. Approximate simplified
model is developed. Ordinary differential equation is derived, and solved numerically,
what gives a possibility to estimate quantitatively the range of parameters where
jets may be stabilized by torsional oscillations. The polytropic equation of state
$P=K\,\rho^\gamma$, with $\gamma=1$, was considered. Using approximate relations

\begin{equation}
\label{eq15}
  \upsilon_r=r\, a(t,z), \quad \upsilon_\varphi=r\, \Omega(t,z),\quad \upsilon_z=0,
 \end{equation}
introduce non-dimensional variables in the plane, where angular velocity $\Omega$ remains zero during oscillations. the variables in this plane are denoted by "tilde".

\begin{equation}
\label{eq48}
\tau=\omega t, \,\, y=\frac{\tilde R}{R_0},\,\, z=\frac{a \tilde R}{a_0 R_0},\,\,
 a_0=\frac{K}{\omega R_0^2}=\omega,\,\,R_0=\frac{\sqrt K}{\omega},
 \end{equation}
Here $\omega$ is the frequency of radial oscillations. In these variables differential equations have a form

\begin{equation}
\label{eq49}
\frac{dy}{d\tau}=z,\quad
\frac{d z}{d\tau}=\frac{1}{y}(1-D\sin^2 \tau);\,\,\,\,y=1,\,\, z=0\,\,{\rm at}\,\, \tau=0.
 \end{equation}
Therefore, the problem is reduced to a system (\ref{eq49}) with only one non-dimensional parameter

$$D=\frac{1}{2\pi K C_m}\left(\frac{C_b\Omega_0}{z_0\omega}\right)^2,$$
where $C_b$ and $C_m$ are integrals of motion, see \cite{bk07} for details.
Solution of this nonlinear system
changes qualitatively with changing of the parameter $D$.
The solution of this system was obtained numerically for $D$ between 1.5, and 3.1. Roughly the solutions may be divided into 3 groups.

1. At $D$ < 2.1  there is no confinement, and radius grows to infinity after several low-amplitude oscillations.

\begin{figure}[h]
\begin{center}
{\includegraphics[width=2.95in]{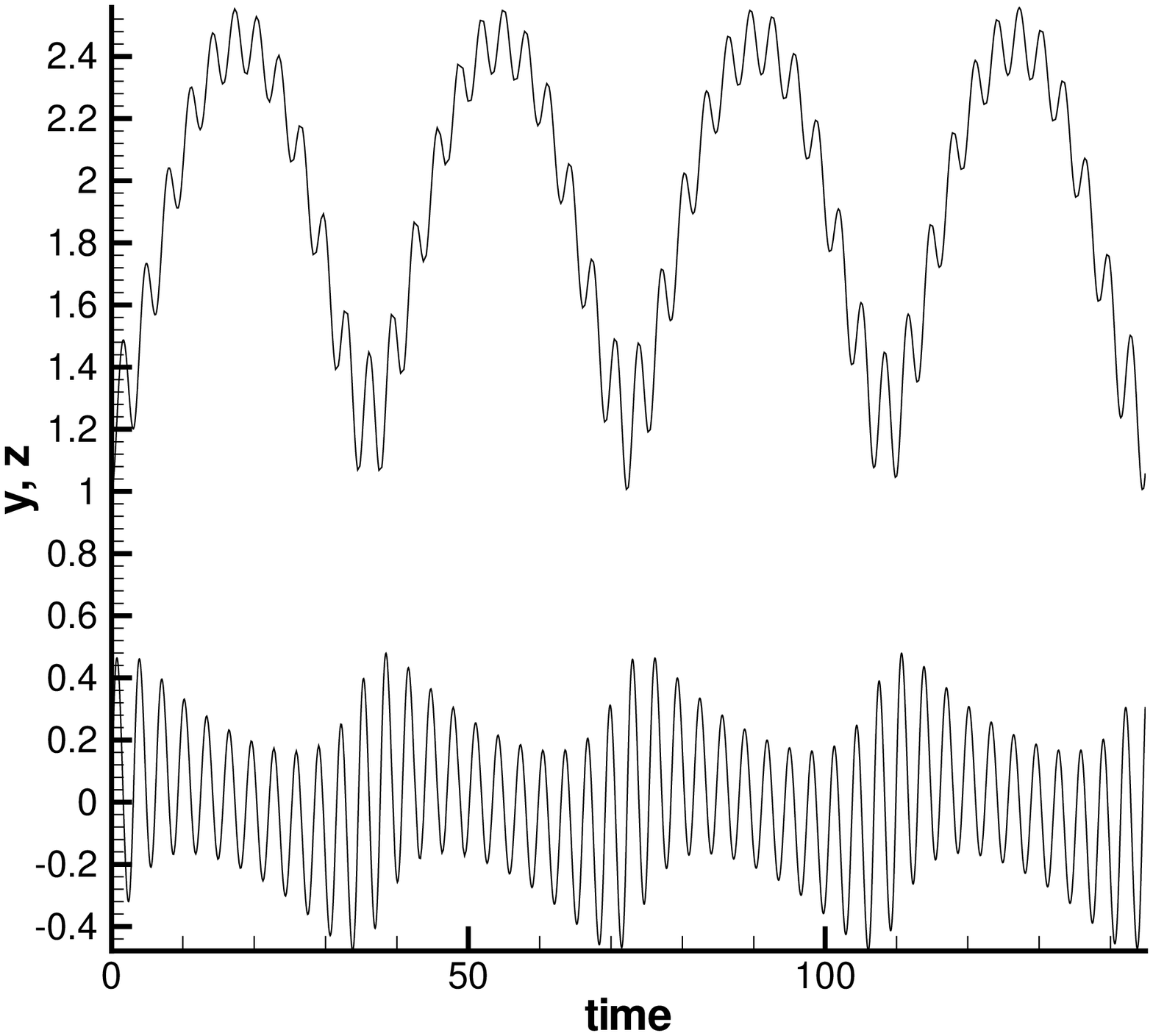}
\includegraphics[width=2.95in]{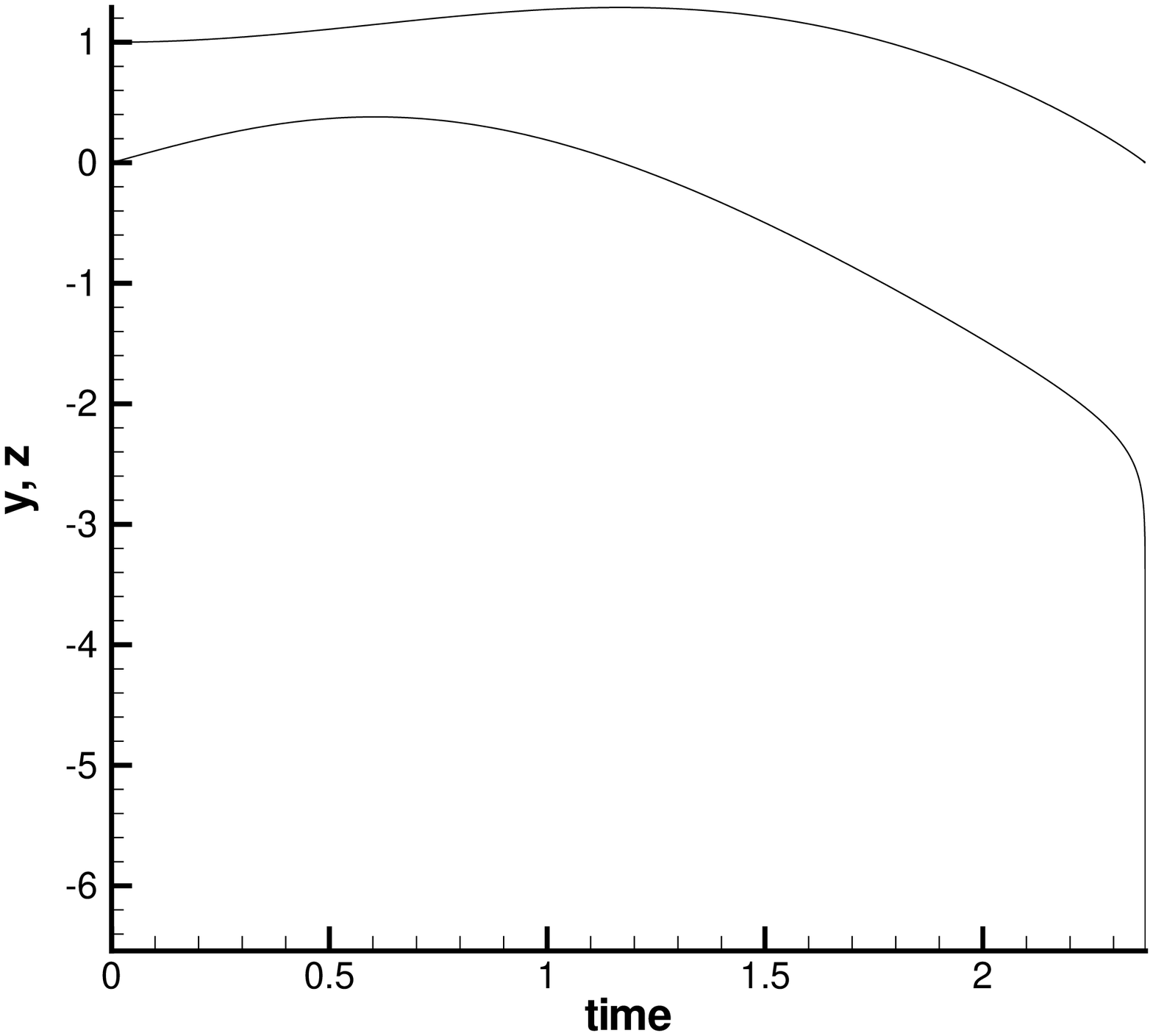}}
\end{center}
\caption{Time dependence of non-dimensional radius $y$ (upper curve), and non-dimensional velocity
$z$ (lower curve), for $D=2.1$ (left), and $D=3.1$ (right).}
\label{fig3}
\end{figure}

2. With growing of $D$, the amplitude of torsional oscillations $\Omega_0$ increases,  and at $D=2.1$ radius is not growing to
infinity, but is oscillating around some average value, forming rather complicated curve (Fig.\ref{fig3}, left).

3. At $D$= 2.28 and larger the radius finally goes to zero with time, forming separate blobs, but with different behavior,
depending on $D$. At $D$ between 2.28 and 2.9 the dependence of the radius $y$ with time may be very complicated,
consisting of low-amplitude and large-amplitude oscillations, which finally lead to zero.
The time at which radius becomes zero depends on $D$ in  rather peculiar way, and may happen at
$\tau \le 100$, like at $D$=2.4, 2.6; or goes trough very large radius, and returned back to zero
value at very large time $\tau \sim 10^7$ at $D$=2.5. From $D=3$ and larger the radius goes to zero at $\tau<2.5$  (Fig.\ref{fig3}, right). before the right side of the second
equation (\ref{eq49}) returned to the positive value. The results of numerical solution are
represented in Figs. 1-18 of \cite{bk07}.
Frequency of oscillations $\omega$ is taken
from linear approximation.
When y(0) is different from 1, there is a larger variety of solutions: regular and chaotic. Development of chaos of these oscillations is analysed in \cite{bk11}

On the edge of the cylinder the rotational velocity cannot exceed the light velocity,
so the solution with initial conditions, corresponding to  $y_0=1$, has
a physical sense only at $\upsilon_s^2 < \frac{c^2}{2\pi^2 D \alpha_n^2} \approx
\frac{c^2}{40 \alpha_n^2}$, where $\omega=\alpha_n\,k\, V_A,\,\, \alpha_n < 1, \,\,
k=\frac{2\pi}{z_0}$, $z_0$ is the space period of the torsional oscillations along $z$ axis, $V_A$ is Alfven velocity. Taking $\alpha_n^2=0.1$ for a strongly non-linear oscillations
we obtain a very moderate restriction $\upsilon_{s0}^2 < \frac{c^2}{4}$. While in the
intermediate collimation regime the outer tangential velocity is not changing significantly,
this restriction would be enough also for the whole period of the time. To have the sound velocity
not exceeding $c/2$, the jet should contain baryons, which density $\rho_0$
cannot be very small, and its
input in the total density in the jet should be larger than about
30\% \cite{bk07}.

\section{Laboratory experiments and numerical simulations}

\begin{figure}[hb]
\begin{center}
\includegraphics[width=2.5in]{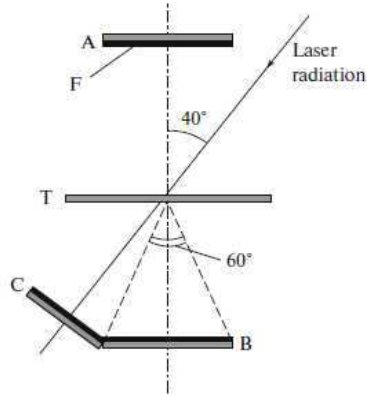}
\end{center}
\caption{Scheme for the experiments investigating the spatial distribution of beams of accelerated protons. A, B, C are the
CR-39 track detectors, F are filters of Al with thicknesses of 11 - 80 $\mu$m, and T is the target of Cu (30 and 50 $\mu$m) or Ta (50 $\mu$m); from \cite{bel18}.}
\label{figexp1}
\end{figure}

Experiments aimed at studying the spatial distributions
of beams of accelerated protons using CR-39
track detectors were carried out at the Neodim 10-
TW picosecond laser facility \cite{bel06}, see Fig.\ref{figexp1}

Numerical simulations of the flow of the matter from the target, heated by the laser beam, have been presented in \cite{bel18}.
Similarity conditions permit to compare parameters of laboratory experiments with jets on the  from AGN's and quasars, see Tabl.

\begin{center}
\begin{tabular}{| l | l |}
\hline
 Laboratory jet &Jets from AGN nuclei (VLBI)  \nonumber\\
 after scaling, from \cite{bel18} &\\\hline
$x=(0.3\div 3)\times 10^{18}$ cm &3$\times 10^{18}$ cm, \\
$t=(0.3\div 3)\times 10^9$ s             & $ 10^{8}$ s, \\
$\upsilon=10^9$ cm/s                 &  $3\times 10^{10}$ cm/s, \\
$\rho=10^{-26}$ g/cm$^3$       &  $10^{-26}$ g/cm$^3$,\\
$n=10^{-2}$ cm$^{-3}$           &   $10^{-2}$ cm$^{-3}$,\\
$H=10^{-1}$ Gs                       &   $10^{-3}$ Gs, \\
$T=10^{11}$ K                       &  $10^{11}$ K.          \\\hline
\end{tabular}
\end{center}

\begin{figure}
\begin{center}
\includegraphics[width=2.2in]{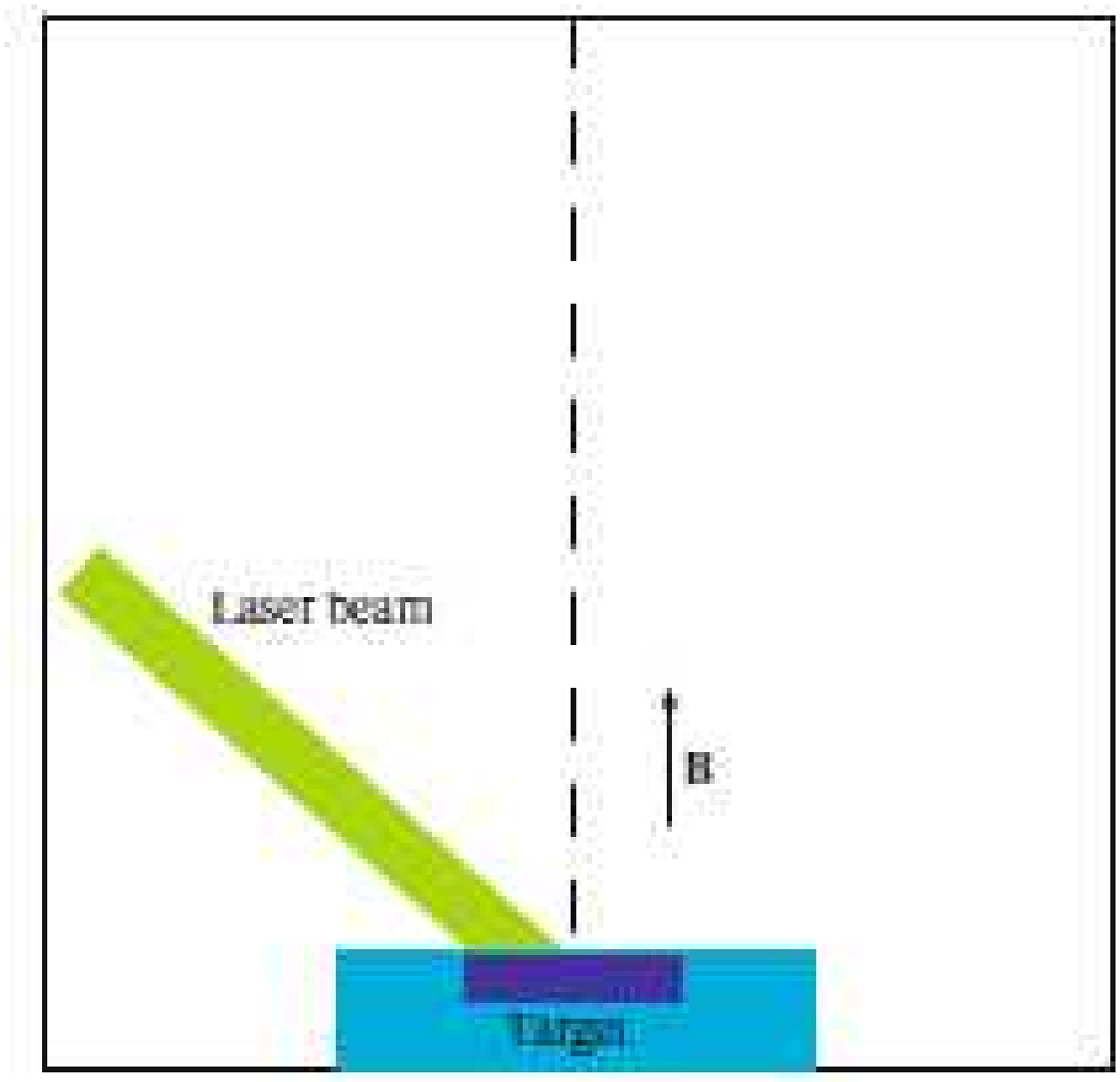}
\includegraphics[width=3.70in]{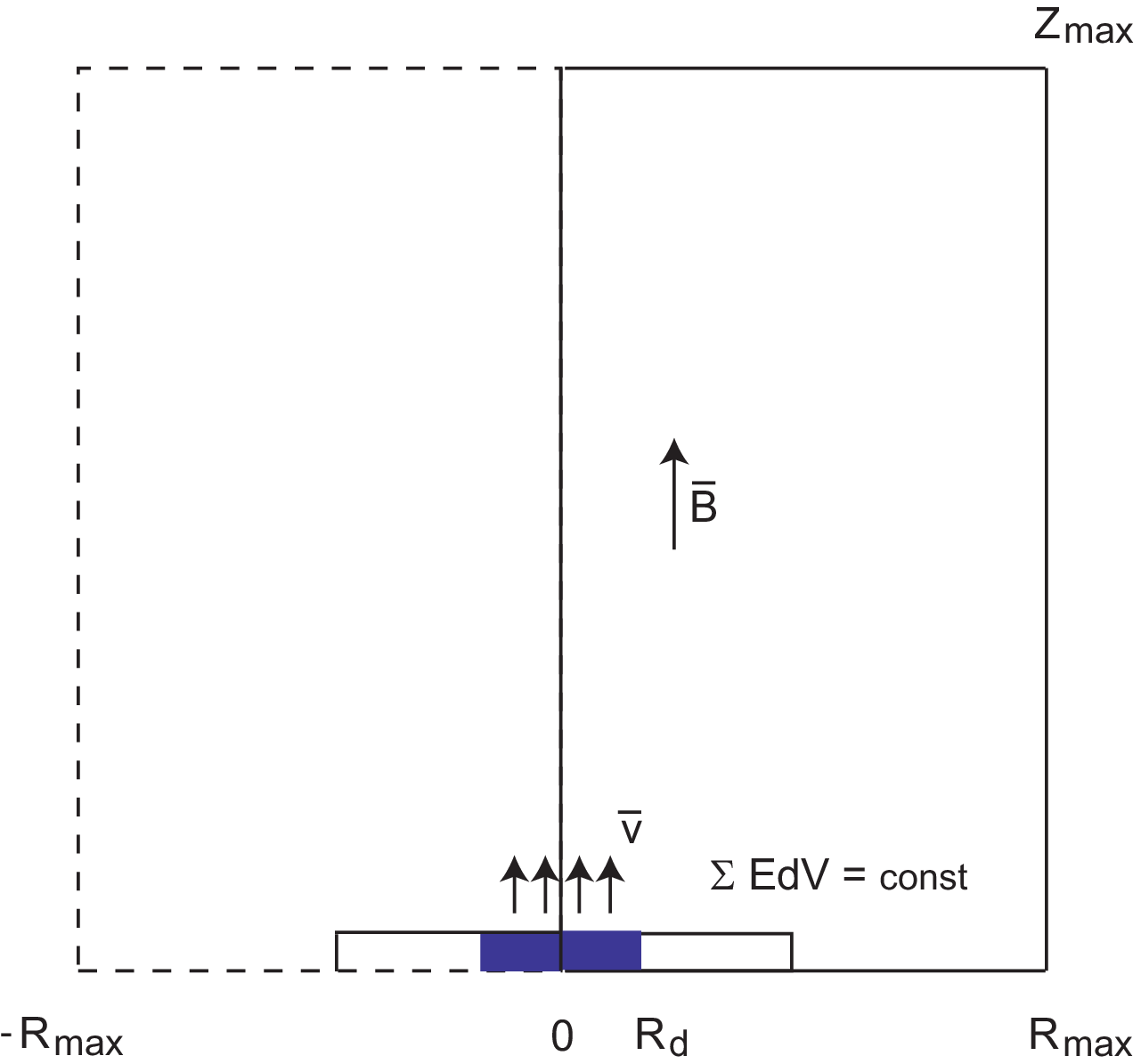}
\end{center}
\caption{Experiment scheme (left). Region of modeling (right), from \cite{bel18}.}
\label{fig1a}
\end{figure}

Analysis of the experiment showed the formation
of a jet due to heating of the foil, with the geometry
of the incident laser beam not being important. This
means that the mathematical model for the formation
of the jet  can be constructed in an axially symmetric
approximation, assuming that the laser-heated spot
is circular.
MHD equations were solved in 2-D problem with a finite electrical conductivity, without gravity. The scheme of the experiment and the region of modeling are presented in Figs.\ref{fig1a}. The detailed description of the method and numerical results are presented in \cite{bel18}.

\begin{figure}[p]
\begin{center}
\includegraphics[width=5.6in]{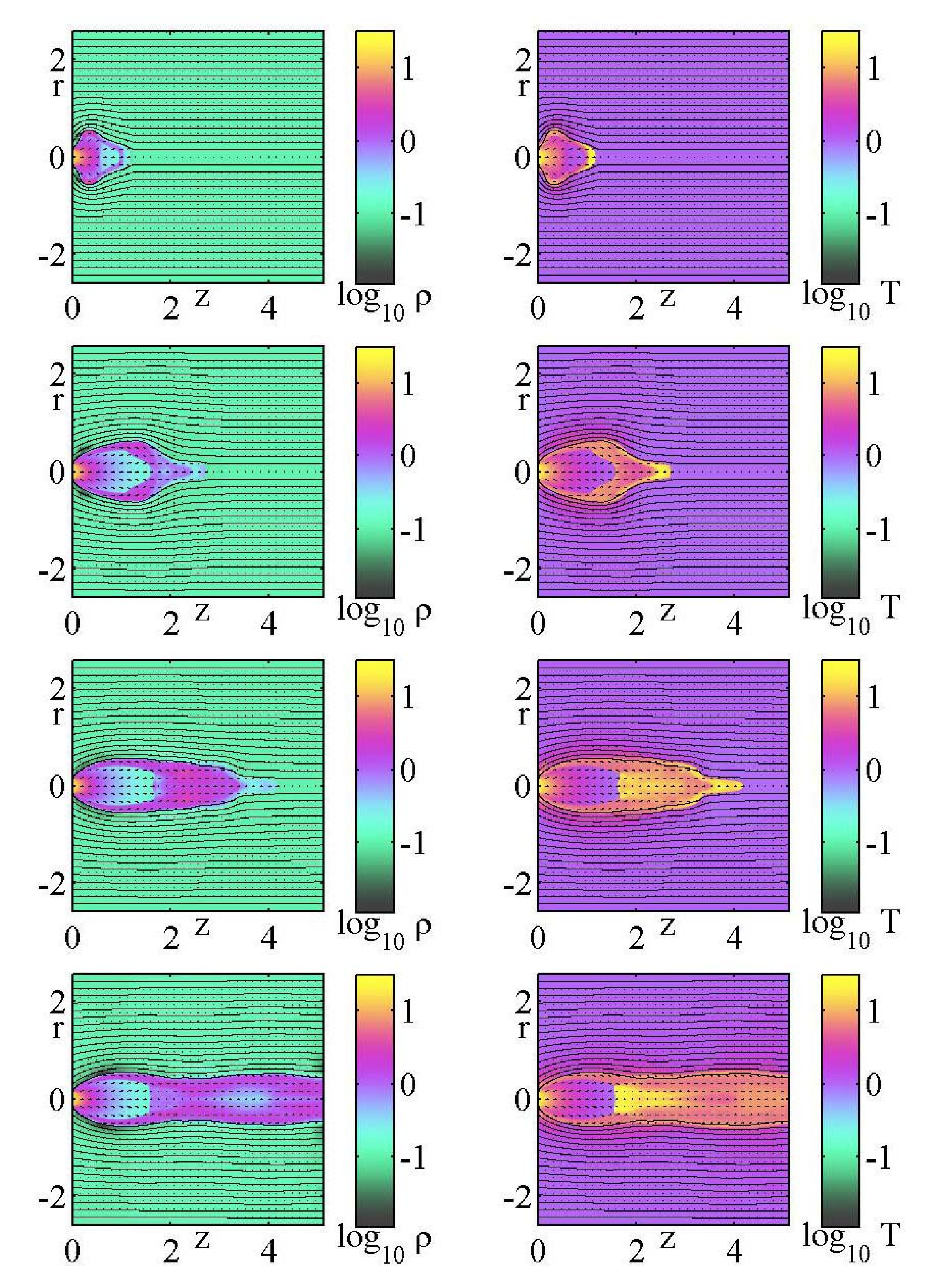}
\end{center}
\caption{Time dependence of the flow, induced by the laser beam, in presence of the poloidal magnetic field with $\beta=0.1$ at time moments t = 5, t = 10, t = 25, t = 45. Density is given in the left column, temperature is to the right, from \cite{bel18}.}
\label{fig9}
\end{figure}

The example of results of numerical calculations,  performed by O.D. Toropina, is presented in Fig.\ref{fig9} from \cite{bel18}, where detailed results of calculations are given.
The ring structure, observed on the photo of the experiment, is visible on this figure. The computations are performed for the target density $\rho\approx 300\rho_0$, where $\rho_0$ is a background density. The important non-dimensional parameter $\beta$ is a ratio of the initial gas pressure at the jet origin to the initial magnetic pressere

\begin{equation}
\label{beta}
\beta =\frac{8\pi P_{0 jet} }{H_{0} ^{2} } =\frac{2}{\gamma } \frac{c_{s0 }^{2} }{V_{A0}^{2} } \; .
\end{equation}

\section{Conclusion - Problems}

1. Jet origin    (blobs or continuous injection; radiation pressure
 or explosions)          BLOBS -!?

 \smallskip

2. Jet collimation (magnetic, or outer pressure, or kinematic)

 \smallskip

3. Jet constitution  (baryonic or pure leptonic)    Baryonic -!?

 \smallskip

4. Particle acceleration (shocks, reconnection, kinetic)

 \smallskip

5. Radiation mechanisms (synchrotron, inverse Compton,
 nuclear processes)

 \smallskip

\indent\indent\indent Jets in Lab should help to answer!

\begin{acknowledgments}
 This work of was partially supported by RFBR grants 17-02-00760, 18-02-00619, and  Fundamental Research Program of Presidium of the RAS \#28.
\end{acknowledgments}

\end{document}